\documentclass[12pt,twoside]{article}
\usepackage[mathscr]{eucal}
\usepackage{amsmath,amsfonts,amssymb,amsthm,mathabx}
\usepackage{times}
\usepackage{pdfsync}
\usepackage{cite}
\usepackage{url}
\usepackage{hyperref}
\usepackage{tensor}
\usepackage{color}
\usepackage{caption}
\usepackage{wrapfig}
\usepackage{float}
\usepackage{graphicx}

\advance\textheight\topskip
\allowdisplaybreaks[1]

\newcommand{\bea}{\begin{eqnarray}}
\newcommand{\eea}{\end{eqnarray}}
\newcommand{\be}{\begin{equation}}
\newcommand{\ee}{\end{equation}}
\newcommand{\bs}{\begin{split}}
\newcommand{\es}{\end{split}}

\newcommand{\poubelle}[1]{}

\newcommand{\dd}{\partial}
\renewcommand{\d}{\partial}

\newcommand{\half}{\frac{1}{2}}

\newcommand{\ffrac}[2]{\raisebox{.5pt}%
  {\footnotesize$\displaystyle\frac{#1}{#2}$}\kern1pt}

\newcommand{\dover}[2]{\ffrac{\dd #1}{\dd #2}}

\newcommand{\ddl}[2]{\ffrac{\dd #1}{\dd #2}}

\def\cD{\mathcal{D}}

\def\cF{\mathcal{F}}

\def\cJ{\mathcal{J}}

\def\cL{\mathcal{L}}
\def\cM{\mathcal{M}}
\def\cN{\mathcal{N}}

\def\cP{\mathcal{P}}

\def\cT{\mathcal{T}}

\def\cY{\mathcal{Y}}

\numberwithin{equation}{section} \makeatletter
\@addtoreset{equation}{section}

\DeclareFontFamily{OT1}{rsfs}{} \DeclareFontShape{OT1}{rsfs}{m}{n}{
<-7> rsfs5 <7-10> rsfs7 <10-> rsfs10}{}
\DeclareMathAlphabet{\mycal}{OT1}{rsfs}{m}{n}
\def\scri{{\mycal I}}%
\def\Scri{\scri}

\newcommand*\xbar[1]{%
  \hbox{%
    \vbox{%
      \hrule height 0.5pt % The actual bar
      \kern0.3ex%         % Distance between bar and symbol
      \hbox{%
        \kern-0.0em%      % Shortening on the left side
        \ensuremath{#1}%
        \kern-0.0em%      % Shortening on the right side
      }%
    }%
  }%
}

\renewcommand{\xbar}{\bar}

\begin{document}

\title{Coadjoint representation of the BMS group on celestial Riemann surfaces}

\author{Glenn Barnich, Romain Ruzziconi}

\date{}

\pagestyle{myheadings} \markboth{\textsc{\small G.~Barnich,
 R.~Ruzziconi}}{\textsc{\small Coadjoint representation of BMS$_4$}}

\addtolength{\headsep}{4pt}

\begin{centering}

  \vspace{1cm}

  \textbf{\Large{Coadjoint representation of the BMS group on
      celestial Riemann surfaces}}

  \vspace{1.5cm}

  {\large Glenn Barnich}

\vspace{.5cm}

\begin{minipage}{.9\textwidth}\small \it \begin{center} Physique
    Th\'eorique et Math\'ematique \\ Universit\'e libre de Bruxelles and
    International Solvay Institutes \\ Campus Plaine C.P. 231, B-1050 Bruxelles,
    Belgium. \\ E-mail: \href{mailto:gbarnich@ulb.ac.be}{gbarnich@ulb.ac.be}
 \end{center}
\end{minipage}

\vspace{.5cm}

{\large Romain Ruzziconi}

\vspace{.5cm}

\begin{minipage}{.9\textwidth}\small \it \begin{center} Institute for
    Theoretical Physics, TU Wien \\ Wiedner Hauptstrasse 8, A-1040 Vienna,
    Austria \\ E-mail:
    \href{mailto:romain.ruzziconi@tuwien.ac.at}{romain.ruzziconi@tuwien.ac.at}
  \end{center}
\end{minipage}

\end{centering}

\vspace{1cm}

\begin{center}
  \begin{minipage}{.9\textwidth} \textsc{Abstract}. The coadjoint
    representation of the BMS group in four dimensions is constructed
    in a formulation that covers both the sphere and the punctured
    plane. The structure constants are worked out for different
    choices of bases. The conserved current algebra of non-radiative
    asymptotically flat spacetimes is explicitly interpreted in these
    terms.
 \end{minipage}
\end{center}

\vfill

\thispagestyle{empty}
\newpage
{\small \tableofcontents}

\vfill
\newpage

\section{Introduction}
\label{sec:introduction}

The BMS group \cite{Bondi:1962px,Sachs1962a,Sachs1962,Penrose1963,Newman1966} is
the symmetry group of four-dimensional asymptotically flat spacetimes at null
infinity. The precise form of this group changes when one replaces the celestial
sphere by a different Riemann surface \cite{Foster1978,Foster1987}. Whereas
unitary irreducible representations of the BMS group are directly relevant for
the quantum theory \cite{Newman:1965fk,Komar1965,McCarthy1975}, the coadjoint
representation is intimately connected to classical solution space through the
momentum map. Unitary irreducible representations come later, after the
classification of coadjoint orbits, via geometric quantization.

In this paper, we provide the detailed construction of the coadjoint
representation of the BMS group and of the algebra on the celestial sphere and
the punctured plane. In the case of the sphere, we explicitly identify the
coadjoint representation in the gravitational data of non-radiative spacetimes.

In order to set the stage, we start in section \ref{Relations with literature}
by providing the (corrected) commutation relations of the $\mathfrak{bms}_4$
algebra on the celestial sphere. Since the structure of the BMS group is the
same as that of the Poincar\'e group in the sense that both are semi-direct
product groups with an abelian ideal, we recall in section
\ref{sec:coadj-repr-semi} the structure of the coadjoint representation of such
groups and algebras \cite{Rawnsley1975}.

In a next step in section \ref{sec:algebra}, we provide a description of the
coadjoint representation of the BMS group in four dimensions in terms of
suitably weighted functions on a two-dimensional surface by focussing on local
aspects. For the presentation, after some generic considerations based on
\cite{Fulton1962,DHoker:1988ta}, we will use a Weyl covariant derivative instead
of the standard ``eth'' operator \cite{Newman1966,Goldberg1967,Held1970}. Note
that our conventions for these derivatives differ somewhat from those originally
introduced in \cite{Geroch:1973am,Penrose:1984,Penrose:1986} for related
reasons. The description applies both to the ``global'' and ``local'' versions
of the algebra
\cite{Barnich:2009se,Barnich:2010eb,Barnich:2011ct,Barnich:2016lyg}, which are
studied explicitly in sections \ref{sec:Concrete realization on the sphere} and
\ref{sec:Concrete realization on the two-punctured Riemann sphere},
respectively. In the former section, we use extensively results of sections 4.14
and 4.15 of \cite{Penrose:1984}. Besides the standard choice of rotation and
boost generators as used originally in \cite{Sachs1962}, we also provide
explicit commutation relations adapted to the $\mathfrak{sl}(2,\mathbb R)\times
\mathfrak{sl}(2,\mathbb R)$ decomposition of the Lorentz algebra. In the latter
section, our expansions follow the standard conventions used in the context of
two-dimensional conformal field theories.

In section \ref{sec:realization-cylinder}, we briefly comment the case of the
cylinder. Finally, in the case of the sphere and for non-radiative
asymptotically flat spacetimes, we explicitly construct the equivariant map from
the free gravitational data at $\Scri^+$ to the coadjoint representation in
section \ref{sec:embedd-into-asympt}.

The coadjoint representation of the BMS group in three dimensions
\cite{Ashtekar1997} (see also \cite{Barnich:2006avcorr,Barnich:2014kra}) has
been investigated in \cite{Barnich:2015uva}. In that case, the abelian factor
can be identified with the Lie algebra of the non abelian factor acted upon by
the adjoint representation, which simplifies the classification of coadjoint
orbits considerably. Furthermore, central extensions are the familiar ones
directly related to the Virasoro group and algebra. Neither of these
simplifications occur in four dimensions. As a consequence, we will not discuss
the classification of coadjoint orbits in this paper. Also, central extensions
that are relevant in the gravitational context are of a different nature
\cite{Barnich:2017ubf}, and will not be considered here.

\section{Poincar\'e and BMS algebras on the celestial sphere}
\label{Relations with literature}

The structure constants of the $\mathfrak{bms}_4$ algebra on the celestial sphere
have been worked out in \cite{Sachs1962} (see also \cite{Antoniou1991} for
corrections, and \cite{Carmeli:2000af}, \cite{Alessio:2017lps} for reviews).
More details on the geometric interpretation can be found in
\cite{Newman1966,Held1970} (see also
\cite{Geroch:1977aa,Penrose:1984,Penrose:1986,Duval:2014uva,Duval:2014lpa}). In
this section, we start by providing the standard commutation relations of the
$\mathfrak{bms}_4$ algebra in terms of rotation and boost generators.

Let $x^a$, $a=0,\dots 3$, be Cartesian coordinates on Minkowski spacetime where
$\eta_{ab}$ = $\text{diag}(1,-1,-1,-1)$ and its inverse $\eta^{ab}$ are used to
lower and raise indices. The starting point is the Poincar\'e algebra with
generators
\begin{equation}
  L^{ab} = L^{[ab]}=-(x^a\dover{}{x_b}-x^b\dover{}{x_a}), \quad P^a=\frac{\partial}{\partial x_a},\label{eq:26}
\end{equation}
satisfying
\begin{equation}
  \begin{split}
[L^{ab}, L^{cd}] &= -(\eta^{bc}
L^{ad} - \eta^{ac} L^{bd} - \eta^{bd} L^{ac} +\eta^{ad} L^{bc}),\\
[P^a,L^{bc}]&=-(\eta^{ab}P^c-\eta^{ac}P^b)
.
\label{Lorentz algebra}
\end{split}
\end{equation}
When splitting into suitable combinations of rotation and boost generators and
of translation generators,
\begin{equation}
  \label{rotboost}
  \begin{split}
    L_z &= L^{12}, \quad 
    L^{\pm} = \pm i L^{23} +L^{13},\quad K_z =
    L^{30}, \quad 
    K^{\pm} = \mp i L^{20} - L^{10},\\
    H &=P^0,\quad P_z=-\frac{1}{2}P^3,\quad P^{\pm}=
 \frac{1}{2}(i P^2 \pm P^1),
  \end{split}
\end{equation}
the non-vanishing commutation relations of the Poincar\'e algebra become
\begin{equation}
  \label{eq:29}
  \begin{split}
    [L^+,L^-]&=2i L_z,\quad [L_z,L^{\pm}]=\pm iL^{\pm}, \quad [K^+,K^-]=-2i L_z,\\
    [K_z,K^{\pm}]&=L^{\pm}, \quad [L^+,K^-]=2 K_z,\quad [L^-,K^+]= 2K_z,\\
    [L_z,K^{\pm}]& =\pm i K^{\pm},\quad [L^{\pm},K_z]=-K^{\pm},\\
    [K_z,H]&=2P_z,\quad [K^\pm,H]=\pm 2 P^\pm,\quad [L^\pm,P_z]=\mp P^\pm,\quad
    [K_z,P_z]=\frac{1}{2}H,\\
    [L_z,P^\pm]&=\pm i P^\pm,\ [L^+,P^-]=-2P_z=-[L^-,P^+],\\
    [K^+,P^-]&=-H=-[K^-,P^+].
\end{split}
\end{equation}

In terms of spherical coordinates and a retarded time coordinate, 
\begin{equation}
  \label{eq:27}
  r=\sqrt{(x^1)^2+(x^2)^2+(x^3)^2},\quad  u=x^0-r,\quad
  r\cos \theta=x^3,\quad r\sin\theta e^{i\phi}=x^1+ix^2, 
\end{equation}
the Poincar\'e generators read 
\begin{equation}
  \begin{split}
    L_z & =\partial_\phi, \\ L^{\pm}   & =- e^{\pm i\phi}\big[\d_\theta\pm i\cot\theta\d_\phi\big],\\
    K_z & =-(1+\frac{u}{r})\cos\theta (r\partial_r)+\cos\theta (u\partial_u)
    +(1+\frac{u}{r})\sin\theta\partial_\theta, \\
    K^{\pm} & =e^{\pm i\phi}\big[(1+\frac{u}{r})\sin\theta (r\partial_r) -
    \sin\theta (u\d_u)+ (1+\frac{u}{r})\cos\theta\d_\theta
    \pm(1+\frac{u}{r})\frac{i}{\sin\theta}\d_\phi
    \big],\\
    H&=\partial_u,\\ -2P_z&=\cos\theta(-\partial_r+\partial_u)
    +\frac{1}{r}\sin\theta\partial_\theta,\\  \pm 2 P^\pm& =e^{\pm
      i\phi}\big[\sin\theta(\partial_r-\partial_u)
    +\frac{1}{r}\cos\theta\partial_\theta\pm\frac{1}{r\sin\theta}\partial_\phi\big].
\end{split}
\end{equation}
As may be shown on general grounds or explicitly checked, the Poincar\'e algebra
in the form \eqref{eq:29} may also be represented in terms of these generators
restricted to the surface $r=cte\to\infty$,
\begin{equation}
  \label{eq:32}
  \begin{split}
    L_z & =\partial_\phi, \quad L^{\pm}  =- e^{\pm i\phi}\big[\d_\theta\pm i\cot\theta\d_\phi\big],\\
    K_z & =\cos\theta (u\partial_u) +\sin\theta\partial_\theta, \quad K^{\pm}
    =e^{\pm i\phi}\big[-\sin\theta (u\d_u)+ \cos\theta\d_\theta
    \pm\frac{i}{\sin\theta}\d_\phi
    \big],\\
    H&=\partial_u,\quad -2P_z=\cos\theta\partial_u,\quad \pm 2 P^\pm =-e^{\pm
      i\phi}\sin\theta\partial_u.
  \end{split}
\end{equation}

The next step is to represent the Poincar\'e algebra at $u=0$. This can be done
by simply restricting the Lorentz generators to that surface, and by
representing the translation generators by suitable functions on that surface,
\begin{equation}
  \label{eq:33}
  \begin{split}
    L_z & =\partial_\phi, \quad L^{\pm}  =- e^{\pm i\phi}\big[\d_\theta\pm i\cot\theta\d_\phi\big],\\
    K_z & =\sin\theta\partial_\theta, \quad K^{\pm} =e^{\pm i\phi}\big[
    \cos\theta\d_\theta
    \pm\frac{i}{\sin\theta}\d_\phi\big],\\
    H&=1={}_0Z_{0,0} ,\quad P_z=-\frac{1}{2}\cos\theta={}_0Z_{1,0},\quad P^\pm
    =\mp\frac{1}{2}e^{\pm i\phi}\sin\theta={}_0Z_{1,\pm 1},
  \end{split}
\end{equation}
while in addition, defining for any function $f$ on the sphere,
\begin{equation}
  \label{eq:34}
  \begin{split}
  [L_z,f]&=L_z(f),\quad [L^\pm,f]=L^\pm (f),\\ [K_z,f]&=K_z(f)-\cos\theta f,
  \quad [K^\pm,f]=K^\pm(f)+e^{\pm i\phi}\sin\theta f.
\end{split}
\end{equation}
When applied to the four functions in the last line of \eqref{eq:33}, this
reproduces the commutation relations of the Lorentz with the translation
generators in the Poincar\'e algebra, i.e., the last two lines of \eqref{eq:29}.
The general expression for the unnormalized spherical harmonics ${}_sZ_{j,m}$
are explicitly given in Appendix \ref{SWSH}.

How the Poincar\'e algebra is enhanced to the $\mathfrak{bms}_4$ algebra in the
context of asymptotically flat spacetimes at null infinity is discussed in the
references at the beginning of this section. Besides the original reference
\cite{Sachs1962}, we also refer to the re-derivation in \cite{Barnich:2010eb}
for more details.

In the $\mathfrak{bms}_4$ algebra, the commutation relations for the Lorentz
sub-algebra are unchanged and given in the first three lines of \eqref{eq:29}.
The commutation relations for the $\mathfrak{bms}_4$ algebra are then completed
by choosing a basis for functions on the sphere. Whatever basis is chosen, the
supertranslation generators $\mathcal T^G_A$ commute,
\begin{equation}
  \label{eq:35}
  [\mathcal T^G_{A},\mathcal T^G_{A'}]=0.
\end{equation}
To get explicit structure constants for the commutators involving Lorentz and
supertranslation generators, one may start with an unnormalized basis involving
associated Legendre functions,
\begin{equation}
  \mathcal T^U_{j,m}=P_j^m(\cos\theta)e^{im\phi},\quad
  P^m_j(x)=(-1)^m(1-x^2)^{\frac{m}{2}}\frac{d^m}{dx^m}P_j(x).
  \label{eq:36}
\end{equation}
The action of the Lorentz generators on the $\mathcal T^U_{j,m}$ is then worked
out according to \eqref{eq:34} by using suitable properties of associated
Legendre functions (see e.g. 8.733 and 8.735 of \cite{GRADSHTEYN1980904}),
\begin{flalign}
  & [L_z, \mathcal T^U_{j,m}] = i m \mathcal T^U_{j,m},
  \label{commutators1nm} \\
 & [L^{+}, \mathcal T^U_{j,m}] = -\mathcal T^U_{j,m+1},  \\
 & [L^{-}, \mathcal T^U_{j,m}] = (j-m+1)(j+m)\mathcal T^U_{j,m-1},
                                 \label{commutators2nm} \\
 & [K_z, \mathcal T^U_{j,m}] = \frac{(j-1)(j-m+1)}{2j+1}\mathcal T^U_{j+1,m}
                               -\frac{{(j+2)(j+m)}}{2j+1} 
                               \mathcal T^U_{j-1,m},
    \label{commutators4nm} \\
 & [K^+, \mathcal T^U_{j,m}] = \frac{j-1}{2j+1}
                               \mathcal T^U_{j+1,m+1}  + \frac{j+2}{2j+1}
                               \mathcal T^U_{j-1,m+1}, 
\label{commutators6nma} 
\end{flalign}
\begin{multline}
[K^-, \mathcal T^U_{j,m}] =  -\frac{(j-1)(j-m+1)(j-m+2)}{2j+1}
\mathcal T^U_{j+1,m- 1}  \\ -
\frac{(j+2)(j+m)(j+m-1)}{2j+1}
                               \mathcal T^U_{j-1,m- 1}. 
                               \label{commutators6nmb}
\end{multline}
For a normalized basis in terms of standard spherical harmonics, 
\begin{equation}
  \label{eq:37}
  \mathcal T^S_{j,m}={}_0Y_{j,m}=\sqrt{\frac{(2l+1)(l-m)!}{4\pi (l+m)!}}
  P_j^m(\cos\theta)e^{im\phi},
\end{equation}
one finds instead
\begin{flalign}
 & [L_z, \mathcal T^S_{j,m}] = i m \mathcal T^S_{j,m} , \label{commutators1}\\
 & [L^{\pm}, \mathcal T^S_{j,m}] = \mp\sqrt{(j\mp m)(j\pm m+1)}
  \mathcal T^S_{j,m\pm 1},\label{commutators2}
\end{flalign}
\begin{multline}
[K_z, \mathcal T^S_{j,m}] = (j-1)
\sqrt{\frac{(j+m+1)(j-m+1)}{(2j+1)(2j+3)}} \mathcal T^S_{j+1,m} \\
-(j+2) \sqrt{\frac{(j+m)(j-m)}{(2j-1)(2j+1)}} \mathcal
T^S_{j-1,m}, \label{commutators4}
\end{multline}
\begin{multline}
[K^\pm, \mathcal T^S_{j,m}] = \pm (j-1) \sqrt{\frac{(j\pm m+2)(j\pm
m+1)}{(2j+1)(2j+3)}} \mathcal T^S_{j+1,m\pm 1} \\ \pm (j+2)
\sqrt{\frac{(j\mp m)(j\mp m-1)}{(2j-1)(2j+1)}} \mathcal T^S_{j-1,m\pm
1}.
\label{commutators6}
\end{multline}
This is the form under which the commutation relations between Lorentz
and supertranslation generators usually appear in the literature (with
due care devoted to various conventions and correction of misprints).

\section{Coadjoint representations of semi-direct product groups and 
  algebras}
\label{sec:coadj-repr-semi}

For a semi-direct product group of the form $G \ltimes_\sigma A$, with
$G$ a Lie group and $A$ an abelian Lie group seen as a vector space
with the addition, the group law is given by 
\begin{equation}
(f, \alpha ) \cdot (g, \beta) = (f\cdot g, \alpha + \sigma_f(\beta)),
\label{produit} 
\end{equation}
while $\sigma$ is a representation of $G$ on $A$. The associated Lie algebra is
$\mathfrak{g} \oright_\Sigma A$, the Lie algebra of $A$ being identified with
$A$ itself, and
\begin{equation}
[(X, \alpha) , (Y,\beta)]= ([X,Y], \Sigma_X \beta -\Sigma_Y \alpha),
\label{commute} 
\end{equation}
where $\Sigma$ is the differential of $\sigma$. The adjoint actions of
the group and the algebra are then given by
\begin{equation}
{\rm Ad}_{(f,\alpha)} (X, \beta) = ({\rm Ad}_f X, \sigma_f \beta -
\Sigma_{{\rm Ad}_f X} \alpha),
\label{Adjointe}
\end{equation}
\begin{equation}
{\rm ad}_{(X, \alpha)} (Y,\beta) = ([X,Y], \Sigma_X \beta - \Sigma_Y
\alpha).
\label{adjointe}
\end{equation}
The dual space to the Lie algebra is given by $\mathfrak{g}^*\oplus
A^*$, with non-degenerate pairing denoted by
\begin{equation}
\langle (j, p), (X, \alpha)\rangle = \langle j, X \rangle + \langle p,
\alpha \rangle,
\label{abstract pairing}
\end{equation}
and coadjoint actions
\begin{align}
\label{eq:79} \langle {\rm Ad}^*_{(f,\alpha)}(j,p),(Y,\beta)\rangle& =
\langle (j,p), {\rm Ad}_{(f,\alpha)^{-1}}(Y,\beta)\rangle,\\ \langle
{\rm ad}^*_{(X,\alpha)}(j,p),(Y,\beta)\rangle& = \langle (j,p), -{\rm
ad}_{(X,\alpha)}(Y,\beta)\rangle.\label{eq:79a}
\end{align}
Defining $\times: A\oplus A^*\to \mathfrak g^* $ by
\begin{equation}
\langle \alpha \times p, X \rangle = \langle p, \Sigma_X \alpha
\rangle,
\end{equation}
and $\sigma^*$ to be the dual representation associated with $\sigma$,
$\sigma^*: G\times A^*\to A^*$,
\begin{equation}
\langle \sigma^*_f p, \alpha \rangle = \langle p, \sigma_{f^-1} \alpha
\rangle,
\end{equation}
the coadjoint representations are given by
\begin{equation}
{\rm Ad}^*_{(f, \alpha)} (j, p) = ({\rm Ad}^*_f j + \alpha \times
\sigma^*_f p, \sigma^*_f p),
\label{CoAdjointe}
\end{equation}
\begin{equation}
{\rm ad}^*_{(X, \alpha)} (j, p) = ({\rm ad}^*_X j + \alpha \times p,
\Sigma^*_X p).
\label{coadjointe}
\end{equation}
In terms of generators, $(e_A,e_\alpha)$ of $\mathfrak g\oright_\Sigma
A$, with $(e_*^A,e_*^\alpha)$ the associated dual basis of
$\mathfrak{g}^*\oplus A^*$,
\begin{equation}
\label{eq:2} [e_A,e_B]=f_{AB}^Ce_C,\quad [e_A,e_\alpha]=
f^\beta_{A\alpha} e_\beta,\quad [e_\alpha,e_\beta]=0,
\end{equation}
the coadjoint representation of the algebra \eqref{coadjointe} becomes
\begin{equation}
\label{eq:3} {\rm ad}^*_{e_A}e_*^B=-f^B_{AC} e_*^C, \quad {\rm
ad}^*_{e_\alpha}e_*^B=0,\quad {\rm
ad}^*_{e_A}e_*^\beta=-f^\beta_{A\gamma} e_*^\gamma,\quad {\rm
ad}^*_{e_\alpha}e_*^\beta=-f^\beta_{\alpha C} e_*^C.
\end{equation}

\section{General structure of the coadjoint representation of BMS4} 
\label{sec:algebra}

\subsection{Background structure}
\label{sec:background-structure}

\subsubsection{Extended conformal transformations}
\label{sec:Extended conformal transformations}

Consider an $n$-dimensional Riemannian manifold with coordinates
$x^\alpha$ and metric $g_{\alpha\beta}(x)$ which transforms under
invertible coordinate transformations $x^{\prime \alpha}=x^{\prime
\alpha}(x)$ as
\begin{equation}
\label{eq:11} g'_{\gamma\delta}(x')
=g_{\alpha\beta}(x)\dover{x^\alpha}{x^{\prime
\gamma}}\dover{x^\beta}{x^{\prime \delta}}.
\end{equation}
Conformal coordinate transformations are such invertible coordinate
transformations for which
\begin{equation}
\label{eq:1} g'_{\gamma\delta}(x')=g_{\gamma\delta}(x')\Omega^2(x').
\end{equation}
Consider then a two-dimensional surface $\mathcal{S}$ with coordinates
$x^\alpha=(\xi,\bar\xi)$ and a conformally flat metric
\begin{equation}
d s^2 = g_{\alpha\beta}dx^\alpha dx^\beta = -2(P\bar P)^{-1}d\xi d\bar\xi,\label{eq:4}
\end{equation}
for some nowhere vanishing $P(x)$. In this case, coordinate
transformations of the form
\begin{equation}
\xi'=\xi'(\xi),\quad \bar\xi'=\bar\xi'(\bar\xi)\label{eq:6}
\end{equation}
are conformal coordinate transformations with
\begin{equation}
\Omega(x')=\left[\frac{(P\bar P)(x')}{(P\bar P)(x)}
J\right]^\half,\quad J=\frac{\d\xi}{\d\xi'}
\frac{\d\bar\xi}{\d\bar\xi'}. \label{eq:70}
\end{equation}
For such conformal transformations, the transformation law
\begin{equation}
\label{eq:84} P'(x')=P(x)\frac{\d\xi'}{\d\xi},\quad \bar P'(x')=\bar
P(x)\frac{\d\bar\xi'}{\d\bar\xi},
\end{equation}
induces the transformation \eqref{eq:1} of the metric components. The
more general transformation law
\begin{equation}
\label{eq:85} P'(x')=P(x)\frac{\d\xi'}{\d\xi}e^{-E(x')},\quad \bar
P'(x')=\bar P(x)\frac{\d\bar\xi'}{\d\bar\xi}e^{-\bar E(x')},
\end{equation}
with $E$ a complex scalar field can be understood as follows. Writing the metric
with suitable zweibeins as,
\begin{equation}
  \label{eq:115}
  ds^2={e^A}_\alpha dx^\alpha \eta_{AB} {e^B}_\beta dx^\beta,\quad \eta_{AB}=\begin{pmatrix}  0 & -1\\ -1& 0 
  \end{pmatrix},\quad {e_1}^\alpha\partial_\alpha=P\partial,\quad {e_2}^\alpha\partial_\alpha=\xbar P\xbar \partial,
\end{equation}
they correspond to the transformations of the zweibeins under conformal
coordinate transformations. At the same time, the imaginary part $iE_I$ produces
a local rotation of the zweibeins while the real part of $E_R$ generates the
Weyl rescaling of the metric. The associated conformal factor is
\begin{equation}
\label{eq:86} \Omega(x')=\left[\frac{(P\bar P)(x')}{(P\bar P)(x)}
J\right]^\half e^{E_R(x')}.
\end{equation}
Three relevant subclasses of the extended transformations
\eqref{eq:85} are

(i) \underline{Conformal coordinate transformation:} taking $E =
0 = \bar{E}$ in \eqref{eq:85} leads back to \eqref{eq:84}, and thus,
when applied to the metric components, to the same transformations
\eqref{eq:1} as those coming from the conformal coordinate
transformation applied to the metric tensor.

(ii) \underline{Complex Weyl rescaling:} taking $\xi'= \xi$,
$\bar\xi'=\bar\xi$ in \eqref{eq:85} gives
\begin{equation}
  \label{Pure Weyl} P'(x)=P(x)e^{-E(x)},\quad \bar P'(x)=\bar
P(x)e^{-\bar E(x)}.
\end{equation}
This induces a real local Weyl rescaling on the metric,
\begin{equation}
  \label{eq:15} g'_{\alpha\beta}(x)=e^{2E_R}g_{\alpha\beta}(x).
\end{equation}
When the real part $E_R=0$, the metric is unchanged. 

(iii) \underline{Fixed conformal factor:} fixing as in
  \cite{Newman1966,Held1970} the conformal factor to be a prescribed function of
  its arguments, $P(x) = P_F(x)$, $\bar P(x)=\bar P_F (x)$ and $P'(x') = P_F
  (x')$, $\bar P'(x')=\bar P_F(x')$ in \eqref{eq:85}, implies that complex Weyl
  rescalings are frozen to
\begin{equation} e^{E(x')} = \frac{P_F(x)}{P_F(x')} \frac{\partial
\xi'}{\partial \xi} \Longleftrightarrow \left\{
\begin{array}{l} e^{E_R(x')} = J^{-\frac{1}{2}} \left[\frac{(P_F
\bar{P}_F)(x)}{(P_F \bar{P}_F) (x')} \right]^\frac{1}{2} , \\
e^{iE_I(x')} = \left[\frac{(P_F/ \bar{P}_F)(x)}{(P_F/ \bar{P}_F)(x')}
\left(\frac{\partial \xi' / \partial \xi }{\partial \bar{\xi}' /
\partial \bar{\xi}} \right) \right]^{\frac{1}{2}} ,
\end{array} \right.
\label{frozen factor expressions}
\end{equation}
where $E = E_R + i E_I$. We will mostly be interested in 2 particular cases
below. The first is when $\mathcal S$ is a 2-sphere of radius $R$ with metric
$ds^2=-R^2(d\theta^2+\sin^2\theta d\phi^2)$ and
$\xi=\bar\zeta=\cot{\frac{\theta}{2}}e^{-i\phi}$, so that
\begin{equation}
\label{eq:12}
P_S=\frac{1+\xi\bar\xi}{R\sqrt 2}=\bar P_S, 
\end{equation}
see \cite{Penrose:1984} section 4.15 for details.

The second is the punctured complex plane, the complex plane with the
origin removed $\mathbb C_0=\mathbb C-\{0\}$, with standard metric
$ {ds}^2=-2dz d\bar z$ so that
\begin{equation}
  \label{eq:13}
  P=1={\bar P}. 
\end{equation}

\subsubsection{Conformal fields and weighted scalars}
\label{sec:Conformal fields and weighted scalars}

Under conformal coordinate transformations and complex Weyl rescalings,
fields $\phi^{\lambda, \bar{\lambda}}_{h,\bar h}$ of conformal
dimensions $(h,\bar h)$ and Weyl weights $(\lambda, \bar{\lambda})$
transform as
\begin{equation}
  \label{eq:93} {\phi'}^{\lambda, \bar{\lambda}}_{h,\bar h} (x')=
e^{\lambda E(x')} e^{\bar{\lambda} \bar{E}(x')} \Big(\ddl{\xi}{\xi'}
\Big)^h \Big(\ddl{\bar\xi}{\bar\xi'}\Big)^{\bar h} \phi_{h,\bar
h}^{\lambda, \bar{\lambda}}(x).
\end{equation}
It follows from \eqref{eq:85} that the conformal dimensions and Weyl
weights of $P$ are both $(-1,0)$ whereas those of $\bar P$ are both
$(0,-1)$. These quantities can be used to map the fields
$\phi^{\lambda, \bar{\lambda}}_{h,\bar h}$ into scalars $\eta^{s,w}$
\begin{equation}
  \label{mapping} \eta^{s,w}=P^h\bar P^{\bar h} \phi_{h,\bar
h}^{\lambda, \bar{\lambda}},
\end{equation}
of spin and conformal weights $[s,w]$, 
\begin{equation}
s=(h-\bar h)- (\lambda - \bar{\lambda}),\quad
w=-(h+\bar h) + (\lambda + \bar{\lambda}),\label{eq:14}
\end{equation}
which transform under conformal coordinate transformations and complex
Weyl rescalings as 
\begin{equation}
  \label{eq:109} \eta^{\prime s,w}(x')=e^{wE_R(x')}e^{-is
E_I(x')}\eta^{s,w}(x).
\end{equation}

\subsubsection{Derivative operators}
\label{sec:Derivative operators}

The only non-vanishing components of the Levi-Civita connection
associated with \eqref{eq:4} are
\begin{equation}
  \label{eq:87} \Gamma^\xi_{\xi\xi}=-\d\ln (P\bar P),\quad
\Gamma^{\bar\xi}_{\bar\xi\bar\xi}=-\bar\d\ln (P\bar P),
\end{equation}
where $\d=\d_\xi,\bar\d=\d_{\bar\xi}$.  Equation \eqref{eq:85} induces
their transformation law under conformal coordinate transformations
combined with Weyl rescalings,
\begin{equation}
  \label{eq:88}
\Gamma^{\prime\xi'}_{\xi'\xi'}(x')=\Gamma^\xi_{\xi\xi}(x)\frac{\d\xi}{\d\xi'}+
\frac{\d\xi'}{\d\xi}\frac{\d^2\xi}{\d\xi'\d\xi'}+2\d'E_R(x'),
\end{equation}
with a similar transformation law for $\Gamma^{\bar\xi}_{\bar\xi\bar\xi}$. In
addition to the conformally flat metric \eqref{eq:4}, one supposes that
$\mathcal{S}$ is endowed with a Weyl connection $({W}, \bar{{W}})$ (see
e.g.~\cite{Fulton1962}) that transforms as
\begin{equation}
  {W}'(x') = \Big( \frac{\partial \xi}{\partial \xi'} \Big)
{W}(x) + 2 \partial ' E_R (x') , \quad \bar{{W}}'(x')
= \Big( \frac{\partial \bar \xi}{\partial \bar \xi'} \Big)
\bar{{W}}(x) + 2 \bar{\partial}' E_R(x').
\end{equation}
Using $P$, $\bar{P}$, ${W}$, $\bar{{W}}$, one can define
\begin{equation}
  \begin{split} &K = \frac{1}{2} (\partial \ln \mu -
    \Gamma^{\xi}_{\xi\xi} ) + {W} = \partial \ln \bar{P} +
    {W} ,\quad O = \frac{1}{2} (\Gamma^\xi_{\xi\xi} -
    \partial \ln \mu ) = - \partial \ln \bar P \quad \\ &\bar K =
    \frac{1}{2} (\bar{\partial} \ln \bar{\mu} -
    \Gamma^{\bar{\xi}}_{\bar{\xi}\bar{\xi}} ) + \bar{{W}} =
    \bar{\partial} \ln P + \bar{{W}} , \quad \bar O =
    \frac{1}{2} (\Gamma^{\bar{\xi}}_{\bar{\xi}\bar{\xi}} - \bar{\partial}
    \ln \bar{\mu} ) = - \bar{\partial} \ln P ,
\end{split} \label{complex weyl connection}
\end{equation}
where $\mu = \frac{\bar{P}}{P}$ is a Beltrami differential. These
objects transform as
\begin{equation}
\begin{split} K'(x') &= \Big( \frac{\partial \xi}{\partial \xi'}
\Big) K (x) + \partial' E(x'), \quad O'(x') = \Big(
\frac{\partial \xi}{\partial \xi'} \Big) O (x) + \partial'
\bar{E}(x'), \\ \bar K'(x') &= \Big( \frac{\partial \bar
\xi}{\partial \bar \xi'} \Big) \bar K (x) + \bar \partial' \bar
E(x'), \quad \bar O'(x') = \Big( \frac{\partial \bar
\xi}{\partial \bar \xi'} \Big) \bar O (x) + \bar \partial'
E(x'). 
\end{split}
\end{equation}
The Weyl covariant derivative can then be defined as
\begin{equation}
\begin{split} D \phi_{h, \bar{h}}^{\lambda, \bar{\lambda}} &= [\nabla
+ (h- \lambda) K + (h - \bar{\lambda} ) O ] \phi^{\lambda,
\bar{\lambda}}_{h, \bar{h}}, \\ \bar{D} \phi_{h, \bar{h}}^{\lambda,
\bar{\lambda}} &= [\bar{\nabla} + (\bar{h} - \bar{\lambda})
\bar K + (\bar{h} - \lambda) \bar O] \phi^{\lambda,
\bar{\lambda}}_{h, \bar{h}} ,
\end{split} \label{D and bar D operators}
\end{equation}
where $\nabla \equiv \nabla_\xi$ and
$\bar{\nabla} \equiv \nabla_{\bar{\xi}}$ are the components of the
covariant derivative associated to the Levi-Civita connection
\eqref{eq:87}. Notice that the field $P$ ($\bar{P}$) of Weyl weights
$(\lambda , \bar{\lambda}) = (-1, 0)$ (resp.
$(\lambda , \bar{\lambda}) = (0, -1)$) and conformal dimensions
$(h, \bar{h}) = (-1,0)$ (resp. $(h, \bar{h}) = (0,-1)$) are
holomorphic (resp. anti-holomorphic) with respect to the Weyl
covariant derivative, namely $\bar{D} P = 0$ (resp. $D \bar{P} =
0$). Under conformal coordinate transformations and complex Weyl
rescalings, we have
\begin{equation}
\begin{split} (D \phi_{h,\bar{h}}^{\lambda, \bar{\lambda}})'(x') &=
e^{\lambda E(x')} e^{\bar{\lambda} \bar{E}(x')} \Big( \frac{\partial
\xi}{\partial \xi'} \Big)^{h+1} \Big( \frac{\partial
\bar{\xi}}{\partial \bar{\xi}'} \Big)^{\bar{h}} (D \phi^{\lambda,
\bar{\lambda}}_{h, \bar{h}}) (x) ,\\ (D \phi^{\lambda,
\bar{\lambda}}_{h,\bar{h}})'(x') &= e^{\lambda E(x')} e^{\bar{\lambda}
\bar{E}(x')} \Big( \frac{\partial \xi}{\partial \xi'} \Big)^h \Big(
\frac{\partial \bar{\xi}}{\partial \bar{\xi}'} \Big)^{\bar{h}+1}
(\bar{D} \phi^{\lambda, \bar{\lambda}}_{h, \bar{h}}) (x) .
\end{split}
\end{equation}
Therefore, the operator $D$ ($\bar{D}$) acts on fields
of Weyl weights $(\lambda, \bar{\lambda})$ and conformal dimensions
$(h, \bar{h})$ to produce fields of Weyl weights $(\lambda,
\bar{\lambda})$ and conformal dimensions $(h+1, \bar{h})$ (resp. $(h,
\bar{h}+1)$).

In the following we will assume that the only fields carrying non-vanishing Weyl
weights are $P,\bar P$. All other fields are thus of the form $\phi_{h,\bar
  h}^{0,0}$ with associated scalars $\eta^{s,w}=P^h\bar P^{\bar h}\phi_{h,\bar
  h}^{0,0}$ so that
\begin{equation}
  \label{eq:17}
  s=h-\bar h,\quad w=-(h+\bar h), \quad h=\frac{s-w}{2},\quad \bar h=-\frac{s+w}{2}. 
\end{equation}
If
\begin{equation}
  \eth \eta^{s,w}=P^{h+1}\bar P^{\bar h} (\nabla
  \phi_{h, \bar{h}}^{0,0}),\quad \bar\eth \eta^{s,w}=P^{h}\bar P^{\bar
    h+1} (\bar{\nabla} \phi_{h,\bar{h}}^{0,0}), \label{eq:16}
\end{equation}
then 
\begin{equation}
\begin{split} \eth \eta^{s,w} = P\bar P^{-s}\d(\bar P^s\eta^{s,w}) =
P (\partial - s O) \eta^{s,w} , \\ \bar \eth \eta^{s,w}
=\bar P P^{s}\bar\d(P^{-s}\eta^{s,w}) = \bar P (\bar \partial + s
\bar O)\eta^{s,w}.
\end{split}
\label{eq:34a}
\end{equation}
in agreement with expressions (4.14.34) and (4.14.33) of
\cite{Penrose:1984}. Under conformal coordinate transformations and
complex Weyl rescalings,
\begin{equation}
  \label{eq:108}
  \begin{split} (\eth \eta^{s,w})'(x')=e^{(w-1)
E_R(x')}e^{-i(s+1)E_I(x')}\big[\eth+(w-s)P\partial
E_R(x'(x))\big]\eta^{s,w}(x), \\ (\bar\eth \eta^{s,w})'(x')=e^{(w-1)
  E_R(x')}e^{-i(s-1)E_I(x')}\big[\bar\eth
+(w+s)\bar P\bar\partial E_R(x'(x))\big]\eta^{s,w}(x).
\end{split}
\end{equation} 
Hence, the scalars $\eth \eta^{s,w}$ and $\bar{\eth} \eta^{s,w}$ transform as
scalars of weights $[s+1,w-1]$ respectively $[s-1,w-1]$ only if $w=s\iff h=0$
respectively $w=-s\iff \bar h=0$. Alternatively, one may limit oneself to
complex Weyl rescalings with $E_R=0$ so that only rotations of the zweibeins are
allowed, with no Weyl rescaling of the metric. In this case, only spin weight
$s$ is relevant.

When using the Weyl covariant derivative $D$ instead of the covariant derivative
$\nabla$ associated to the Christoffel connection, this issue does not arise.
Denoting $\mathcal{D} \eta^{s,w}$ and $\bar{\mathcal{D}} \eta^{s,w}$ the images
under the mapping \eqref{mapping} of $D \phi_{h, \bar{h}}^{\lambda,
  \bar{\lambda}}$ and of $\bar{D} \phi_{h, \bar{h}}^{\lambda, \bar{\lambda}}$,
respectively, we have
\begin{equation}
  \begin{split}
\mathcal{D} \eta^{s,w} &= \left[ \eth +
\left(\frac{s-w}{2} \right) (\mathcal{O} + \mathcal{K}) \right]
\eta^{s,w} = P\left[\partial - \frac{s}{2}( O -
K) - \frac{w}{2} (O + K)
\right]\eta^{s,w},\\ \bar{\mathcal{D}} \eta^{s,w} &=\left[ \bar{\eth}
- \left(\frac{w+s}{2} \right) (\bar{\mathcal{O}} + \bar{\mathcal{K}})
\right] \eta^{s,w} = \bar P\left[\bar \partial + \frac{s}{2}(
\bar O  - \bar K ) - \frac{w}{2}
(\bar O + \bar K ) \right]\eta^{s,w}, 
\end{split}
\label{New operators}
\end{equation}
where $\mathcal{O}= PO$, $\mathcal{K}=PK$,
$\bar{\mathcal{O}}=\bar P\bar O$,
$\bar{\mathcal{K}}=\bar P\bar K$.  Under conformal coordinate
transformations and complex Weyl rescalings, we now have
\begin{equation}
\begin{split} (\mathcal{D} \eta^{s,w})'(x') &= e^{(w-1) E_R(x')} e^{-i
(s+1) E_I(x')} (\mathcal{D} \eta^{s,w})(x), \\ (\bar{\mathcal{D}}
\eta^{s,w})'(x') &= e^{(w-1) E_R(x')} e^{-i (s-1) E_I(x')}
(\bar{\mathcal{D}} \eta^{s,w})(x) .
\end{split}
\label{transfo new op}
\end{equation} Therefore, the operator $\mathcal{D}$
($\bar{\mathcal{D}}$) acts on weighted scalars $[s,w]$ to produce
weighted scalars $[s+1, w-1]$ (resp. $[s-1,w-1]$).
The following property holds:
\begin{equation}
  [\mathcal{D},\bar{\mathcal{D}}] \eta^{s,w} = -P\bar P\big(s
  \partial \bar{\partial} \ln (P \bar{P}) 
  +\frac{s-w}{2}\bar\partial W +\frac{s+w}{2} \partial \bar
  W \big)\eta^{s,w}.
\end{equation}
Note that $R =-2 P \bar{P} \partial \bar{\partial} \ln (P \bar{P})$ is
the scalar curvature of $\mathcal S$.

\subsubsection{Ingredients}
\label{sec:Ingredients}

In the considerations below, all fields except for $P,\bar P$ have
Weyl weights $(0,0)$. We need the following ingredients:

(i) \underline{Supertranslation field:} a real conformal field
$\tilde{\cT}$ of dimensions $(-\half,-\half)$ and its assocated weighted scalar
$\cT$ under the map \eqref{mapping} of weights $[0,1]$.

(ii) \underline{Superrotation field:} a conformal field
$\tilde{\cY}$ of dimensions $(-1,0)$, its associated weighted scalar
$\cY$ of weights $[-1,1]$, and the complex conjugates
$\bar{\tilde{\cY}}$ and $\bar{\cY}$. These fields satisfy the
conformal Killing equation which becomes
\begin{equation}
  \label{eq:25} \bar{\mathcal{D}} \cY=0\iff \bar D \tilde{\cY}=0,
\end{equation}
together with the complex conjugate relations. Locally, the solutions are simply
$\tilde{\cY}=\tilde{\cY}(\xi)$ and $\cY=P^{-1}\tilde{\cY}(\xi)$, with
$\tilde{\cY}(\xi)$ arbitrary. This will not be the case when taking global
restrictions into account. Note also that, because $s=-w$ for $\cY$, it follows
from the second of \eqref{New operators} that the first of \eqref{eq:25} can
also be written using $\bar\eth$ instead of $\bar{\mathcal{D}}$.

(iii) \underline{Supermomentum:} a real conformal field
$\tilde{\cP}$ of dimensions $(\frac{3}{2},\frac{3}{2})$ and its
associated weighted scalar $\cP$ of weights $[0,-3]$.

(iv) \underline{Super angular momentum:} a conformal field
$\tilde{\cJ}$ of dimensions $(1,2)$ and its associated weighted scalar $\cJ$ of
weights $[-1,-3]$, together with the complex conjugates $\bar{\tilde{\cJ}}$ and
$\bar{\cJ}$. We consider equivalence classes $[\cJ]$ such that $\cJ\sim
\cJ+\mathcal{D} \cL$ with $\cL$ characterized by the weights $[-2,-2]$ and their
complex conjugates. In this case, it follows from the first of \eqref{New
  operators} that, since $s=w$ for $\cL$, one may also write $\eth \cL$ in the
equivalence relation. Similarly, we consider equivalence classes $[\tilde{\cJ}]$
such that $\tilde{\cJ} \sim \tilde{\cJ}+D \tilde{\cL}$ with $\tilde{\cL}$
characterized by the conformal dimensions $(0,2)$ and their complex conjugates).
These equivalence classes may be called super angular momenta.

The conformal dimensions, the Weyl weights, and the spin and conformal weights
of the different ingredients are summarized in the tables \ref{weights S} below.
The objects $\tilde{d\mu}$, $d\mu$ represent the integration measure (see
below).

\begin{table}[h]
\begin{center}
  \begin{tabular}{ | c | c | c | c | c | c | c | c|} \hline $\phi_{h,\bar{h}}$ & $
    \tilde{\cT}$ & $ \tilde{\cY} $ & $ \tilde{\cP} $ & $ \tilde{\cJ} $ &
    $\tilde{d\mu}$ & $P$ & $D$ \\ \hline $h$ & $-\frac{1}{2}$ & $-1 $ &
    $\frac{3}{2}$ & $1$ & $-1$ & $-1$ & $1$ \\ \hline $\bar h$ & $-\frac{1}{2}$ & $0$
    & $\frac{3}{2}$ & $2$ & $-1$ & $0$ & $0$ \\ \hline $\lambda$ & $0$ & $0$ & $0$ &
    $0$ & $0$ & $-1$ & $0$\\ \hline $\bar{\lambda}$ & $0$ & $0$ & $0$ & $0$ & $0$ &
    $0$ & $0$\\ \hline
  \end{tabular} $~~~~$
  \begin{tabular}{ | c | c | c | c | c | c | c|} \hline $\eta^{s,w}$ &
$\cT$ & $\cY$ & $\cP$ & $\cJ$ & $d\mu$ & $\cD $\\ \hline $s$ & $0$ & $-1$ &
$0$ & $-1$ & $0$ & $1$ \\ \hline $w$ & $1$ & $1$ & $-3$ & $-3$ & $2$ & $-1$ \\
\hline
  \end{tabular} \caption{Dimensions and weights} \label{weights S}
\end{center}
\end{table}
Under complex conjugation, 
$\overline{(h, \bar{h})} = ( \bar{h} , h)$,
$\overline{(\lambda , \bar{\lambda})} = (\bar{\lambda} , \lambda )$,
$\overline{[s,w]} = [-s,w]$.

\subsection{Coadjoint representation of the algebra}
\label{sec:algebra-1}

\subsubsection{Weighted scalars}

In terms of above ingredients, the $\mathfrak{bms}_4$ algebra may be
defined by triplets $s=(\cY,\bar\cY,\cT)$ with the commutation
relations
\begin{equation} [(\cY_1, \bar{\cY}_1, \cT_1), (\cY_2, \bar{\cY}_2,
\cT_2) ] = (\hat{\cY}, \hat{\bar{\cY}}, \hat{\cT}), \label{algebrabms}
\end{equation} where
\begin{equation} \left\{
      \begin{aligned} \hat{\cY} &= \cY_1 \mathcal{D} \cY_2 - \cY_2
\mathcal{D} \cY_1\,, \\ \hat{\cT} &= \cY_1 \mathcal{D} \cT_2 -\half
\mathcal{D} \cY_1 \cT_2 -(1\leftrightarrow 2) +{\rm c.c.}\,. \\
      \end{aligned} \right.
    \label{commutator0}
  \end{equation} Elements of the type $(\cY,\bar\cY,0)$ form a
  sub-algebra $\mathfrak g$.  As usual, we identify the individual
  entries of the triplets/doublets with the triplets/doublets where all
  other entries are zero.  For weighted scalar $\eta^{s,w}$, a representation of
  $\mathfrak g$ is defined by 
\begin{equation}
  \label{eq:49} \cY\cdot \eta^{s,w} =\cY \mathcal{D} \eta^{s,w}
  +\frac{s-w}{2} \mathcal{D}\cY \eta^{s,w},\quad
  \bar\cY\cdot \eta^{s,w}=\bar\cY \bar{\mathcal{D}} \eta^{s,w}
  -\frac{s+w}{2} \bar{\mathcal{D}}\bar\cY \eta^{s,w}.
\end{equation}
At this stage, we note that this representation, and also the $\mathfrak{bms}_4$
algebra above and the coadjoint representation below, may also be written with
$\eth,\bar\eth$ instead of $\mathcal D,\bar{\mathcal D}$ because the additional
terms cancel.

In the notation of section \ref{sec:coadj-repr-semi}, we thus have
$X=(\cY,\bar\cY)$, $\alpha=\cT$, and
\begin{equation} \Sigma_X\alpha=(\cY,\bar\cY)\cdot \cT=\cY
\mathcal{D} \cT-\half \mathcal{D} \cY \cT +{\rm c.c.}\,\label{eq:7}.
\end{equation}

Elements of $\mathfrak{bms}^*_4$ are denoted by triplets $([\cJ],[\bar
\cJ],\cP)$ where the pairing is given by
\begin{equation}
\langle([\cJ],[\bar \cJ], \cP), (\cY, \bar{\cY}, \cT)
\rangle = \int_{\mathcal{S}} d\mu\, [\bar \cJ \cY+
\cJ\bar\cY+\cP \cT]. \label{eq:312}
\end{equation}
The measure 
\begin{equation}
  d\mu(\xi,\bar\xi)=\frac{iC}{P\bar P}d\xi\wedge d\bar\xi, \label{eq:18}
\end{equation}
for some normalization constant $C$, has dimensions $(0,0)$ and weights $[0,2]$.
At this stage, we assume that the integral annihilates total $\mathcal{D}$ and
$\bar{\mathcal{D}}$ derivatives. Furthermore, we require the pairing to be
non-degenerate, which can only be the case when taking quotients with respect to
the equivalence relations ${\cJ} \sim {\cJ} + \mathcal{D} {\mathcal{L}}$ and
${\bar{\cJ}} \sim {\bar{\cJ}} + \bar{\mathcal{D}} {\bar{\mathcal{L}}}$. Concrete
realizations where these assumptions hold will be discussed below.

From the definition of the coadjoint representation \eqref{eq:79a}, it then
follows that
\begin{equation}
\begin{split} {\rm ad}^*_{(\cY, \bar{\cY}, \cT)}\cJ& = \bar{\cY}
\bar{\mathcal{D}} \cJ+ 2 \bar{\mathcal{D}} \bar{\cY} \cJ +\mathcal{D}(
\cY \cJ) + \frac{1}{2}\cT \bar{\mathcal{D}} \cP+ \frac{3}{2} \bar{\mathcal{D}} \cT \cP ,\\
{\rm ad}^*_{(\cY, \bar{\cY}, \cT)}\cP &= \cY \mathcal{D} \cP +
\frac{3}{2}\mathcal{D} \cY \cP +{\rm c.c.}\,,
\end{split}
\label{explicit coadjoint}
\end{equation}
where the third term in the first of the above equations does not appear but can
be added because it is equivalent to zero. This is useful in order to have a
transformation law consistent with the conformal dimensions of $\cJ$.

{\bf Remarks:}

(i) The definition makes sense on the level of equivalence classes, 
\begin{equation}
{\rm
  ad}^*_{(\cY, \bar{\cY}, \cT)}([0],[0],0)=([0],[0],0),\label{eq:121}
\end{equation}
because
\begin{equation}
  \bar{\cY} \bar{\mathcal{D}} \mathcal{D} \cL+ 2
\bar{\mathcal{D}} \bar{\cY} \mathcal{D} \cL
=\mathcal{D}(\bar\cY\bar{\mathcal{D}} \cL+2\bar{\mathcal{D}}\bar\cY
\cL).
\end{equation}

(ii) Since the inner product involves complex conjugation, we have
\begin{equation}
  \label{eq:66}
  {\rm ad}^*_{\cY}\cJ= \bar{\cY}
  \bar{\mathcal{D}} \cJ+ 2 \bar{\mathcal{D}} \bar{\cY} \cJ, \quad  {\rm ad}^*_{\bar \cY}\cJ= \mathcal{D}(
  \cY \cJ)\sim 0,\quad {\rm ad}^*_{\cY} \cP = \bar\cY \bar{\mathcal{D}} \cP +
  \frac{3}{2}\bar{\mathcal{D}} \bar\cY \cP, 
\end{equation}
together with the complex conjugates of these relations.

(iii) In the notation of section \ref{sec:coadj-repr-semi},
$j=([\cJ],[\bar\cJ])$, $p=\cP$ and
\begin{equation}
  \label{eq:8}
  \begin{split} \Sigma^*_Xp &=\cY \mathcal{D} \cP +
\frac{3}{2}\mathcal{D} \cY \cP +{\rm c.c.}\,,\\ \alpha\times p &=([\frac{1}{2}\cT
\bar{\mathcal{D}} \cP+\frac{3}{2}\bar{\mathcal{D}} \cT \cP],[\frac{1}{2}\cT \mathcal{D}
\cP+\frac{3}{2}\mathcal{D} \cT \cP])\,.
\end{split}
\end{equation}
This relation encodes the change of super angular momentum under an
infinitesimal supertranslation, which depends linearly on supermomentum. Note
also that when using a vector $\cT$ which is not real, the contribution of
$\alpha\times p$ to $\cJ$ is $\frac{1}{2}\bar \cT
\bar{\mathcal{D}} \cP+\frac{3}{2}\bar{\mathcal{D}} \bar \cT \cP$. 

(iv) On the level of integrands, if we define 
\begin{equation}
  \label{eq:5} \cJ^u_s=\bar \cJ\cY+ \cJ\bar\cY+\cP\cT,
\end{equation}
equation \eqref{eq:79a} reads
\begin{equation}
  \label{eq:9} {\rm ad}^*_{s_1}
  \cJ^u_{s_2}=-\cJ^u_{[s_1,s_2]}+\mathcal{D}\cL_{s_1,s_2}
  +\bar{\mathcal{D}}\bar\cL_{s_1,s_2},
\end{equation}
where
\begin{equation}
  \label{eq:10} \cL_{s_1,s_2}=\bar
  \cJ\cY_1\cY_2+(\cJ\cY_1\bar\cY_2)+\cP\cY_1\cT_2+\frac{1}{2}\cP\cT_1\cY_2.
\end{equation}

(v) In the case when there is no non-degenerate pairing, it is still true that
the vector space of elements $([\cJ],[\bar \cJ],\cP)$ forms a representation
under $\mathfrak{bms}_4$. This representation $\rho_{(\cY,\bar \cY,\cT)}$, which
is no longer the coadjoint representation, is defined by replacing ${\rm
  ad}^*_{(\cY,\bar \cY,\cT)}$ by $\rho_{(\cY,\bar \cY,\cT)}$ in the left hand
side of \eqref{explicit coadjoint}.

\subsubsection{Conformal fields}

The above definitions are expressed in terms of weighted scalars. The
analogous definitions in terms of the associated conformal fields are
obtained by a direct rewriting that consists in adding tilde's on all
the scalars and replacing $\mathcal{D}$ by $D$ and $\bar{\mathcal D}$
by $\bar D$.

In particular, under the mapping \eqref{mapping}, the representation
\eqref{eq:49} becomes
\begin{equation}
  \label{eq:49a} \tilde \cY\cdot \phi_{h,\bar h} =\tilde \cY {D} \phi_{h,\bar h}
  +h {D}\tilde \cY \phi_{h,\bar h},\quad
  \tilde{\bar\cY}\cdot \phi_{h,\bar h}=\tilde{\bar\cY} \bar{{D}} \phi_{h,\bar h}
  +\bar h \bar{D}\tilde{\bar\cY} \phi_{h,\bar h}.
\end{equation}
The Weyl covariant derivatives $D,\bar D$ may again be replaced by the ordinary
derivatives $\partial,\bar\partial$ in these representations because the
additional terms cancel.

The pairing is given by
\begin{equation} \langle([\tilde{\cJ}],[\tilde{\bar\cJ}],
  \tilde{\cP}), (\tilde{\cY}, \tilde{\bar{\cY}}, \tilde{\cT})
\tilde \rangle =\int_{\mathcal{S}} \tilde{d\mu}\, [\tilde{\bar\cJ}
\tilde{\cY}+ \tilde{\cJ} \tilde{\bar\cY} + \tilde{\cP}
\tilde{\cT}], \label{eq:312bis}
\end{equation}
where the measure has conformal dimensions $(-1,-1)$, Weyl weights $(0,0)$ and
is associated to the measure \eqref{eq:18} through $d\mu = (P\bar{P})^{-1}
\tilde{d\mu}$, so that
\begin{equation}
  \label{eq:19}
  \tilde{d\mu}=i C d\xi\wedge d\bar\xi. 
\end{equation}

\subsection{Coadjoint representation of the group}
\label{sec:group}

\subsubsection{Conformal fields}
\label{sec:Conformal fields}

Consider conformal coordinate transformations, $(\xi'(\xi),
\bar{\xi}'(\bar{\xi}))=(g(\xi),\bar g(\bar\xi))$, such that $\frac{\partial
  g}{\partial \xi}>0$, $\frac{\partial \bar g}{\partial \bar \xi}>0$. They form
a group $G$ under composition. For a conformal field ${\phi_{h,\bar h}}(x)$ of dimensions
$(h,\bar h)$ (and vanishing Weyl weights), a representation of $G$ is defined
through
\begin{equation}
  \label{eq:75a} \big((g,\bar g)\cdot
{\phi}_{h,\bar h}\big)(x')=\big(\frac{ \partial \xi}{ \partial
\xi'} \big)^{h}\big(\frac{ \partial \bar \xi}{ \partial \bar \xi'}
\big)^{\bar h}{\phi}_{h,\bar h}(x).
\end{equation}

The BMS$_4$ group is determined by elements $(g,\bar g,\tilde{\cT})$ with
multiplication
\begin{equation}
  \label{eq:74} (g_1,\bar{g}_1,\tilde{\cT}_1)\cdot(g_2,\bar
g_2,\tilde{\cT}_2)=\big(g_1\circ g_2,\bar g_1\circ \bar
g_2,\tilde{\cT}_1+(g_1,\bar g_1) \cdot \tilde{\cT}_2\big).
\end{equation}
Elements of the form $(g,\bar g,0)$ form a subgroup isomorphic to $G$. In the
notation of section \ref{sec:coadj-repr-semi}, we thus have $f=(g,\bar g)$,
$X=(\tilde\cY, \tilde{\bar{\cY}})$ and $\alpha=\tilde\cT$ with
\begin{equation}
  \label{eq:71a} \big(\sigma_f(\alpha)\big)(x')=\big((g,\bar g)\cdot \tilde{\cT}\big)(x')
  = \big(\frac{ \partial \xi}{ \partial \xi'}\big)^{-\half} \big(\frac{ \partial \bar \xi}{\partial \bar \xi'}
  \big)^{-\half}\tilde{\cT}(x) .
\end{equation}
For the adjoint action, defined in equation \eqref{Adjointe}, we get
\begin{equation}
  \label{eq:77} \big({\rm Ad}_f X\big)(x')=\big((g,\bar g)\cdot
(\cY,\bar\cY)\big)(x')=\Big(\big(\frac{ \partial \xi}{ \partial \xi'}
\big)^{-1} \tilde{\cY}, \big(\frac{\partial \bar \xi}{
\partial \bar \xi'}\big)^{-1}\tilde{\bar{\cY}}\Big)(x),
\end{equation}
\begin{equation}
  \label{eq:78} \big(\Sigma_{{\rm Ad}_f X}\alpha\big)(x')=\big(\frac{\partial
    \xi}{\partial \xi'}\big)^{-\frac{1}{2}}
  \big(\frac{\partial\bar{\xi}}{\partial\bar{\xi}'}\big)^{-\frac{1}{2}}
  (\tilde{\mathcal{Y}}D \tilde{\mathcal{T}} -
  \frac{1}{2} D
  \tilde{\mathcal{Y}} \tilde{\mathcal{T}}  + {\rm c.c.})(x),
\end{equation}
whereas definition 
\eqref{CoAdjointe} for the coadjoint representation gives 
\begin{equation}
  \big({\rm Ad}^*_{f}\tilde{\mathcal{J}}\big)(x') =
\big(\frac{\partial \xi}{\partial \xi'}\big) \big(\frac{\partial
\bar{\xi}}{\partial \bar{\xi}'}\big)^2 \tilde{\mathcal{J}}(x) , \label{eq1a}
\end{equation}
\begin{equation}
  \big(\sigma^*_{f}\tilde{\mathcal{P}}\big)(x') =
\big(\frac{\partial \xi}{\partial \xi'}\big)^{\frac{3}{2}}
\big(\frac{\partial \bar{\xi}}{\partial
\bar{\xi}'}\big)^{\frac{3}{2}} \tilde{\mathcal{P}}(x)
, \label{eq2a}
\end{equation}
\begin{equation}
  \big(\tilde{\mathcal{T}} \times \sigma^*_{f}
\tilde{\mathcal{P}}\big)(x') = \Big( \big(\frac{\partial
\xi}{\partial \xi'}\big) \big(\frac{\partial \bar{\xi}}{\partial
\bar{\xi}'}\big)^2 (\frac{1}{2}\tilde{\mathcal{T}}\bar{D} \tilde{\mathcal{P}}
+\frac{3}{2} \bar D
\tilde{\mathcal{T}}\tilde{\mathcal{P}}),{\rm c.c.}
\Big)(x), \label{eq3}
\end{equation}
where ${\rm c.c.}$ denotes the complex conjugate of the expression to the left
of the comma.

{\bf Remarks:}

(i) As usual for diffeomorphisms, the adjoint and coadjoint representations of
the algebra discussed previously are the differentials of those of the group
discussed in this section up to an overall minus sign.

(ii) As usual in conformal field theory, on the level of the algebra, we will
consider not only the Lie algebra of the globally well-defined conformal
transformations but also the algebra of infinitesimal local conformal
transformations.

(iii) The formulas for the group can also be used to understand how the coadjoint
representation behaves under conformal mappings. We briefly discuss the
standard map from the punctured plane to the cylinder below. 

\subsubsection{Weighted scalars}
\label{sec:Weighted scalars}

The description in terms of weighted scalars is very similar. As compared to the
previous section, one simply removes the tilde's and replaces $D$, $\bar D$ by
$\mathcal D$, $\bar{\mathcal{D}}$, while at the same time replacing
$(\frac{\partial\xi}{\partial \xi'})^h(\frac{\partial\bar \xi}{\partial \bar
  \xi'})^{\bar h}$ by $e^{w E_R(x')} e^{-is E_I(x')}$ using table \ref{weights S}.

For future reference, let us nevertheless provide explicit formulas. In this
case, the BMS$_4$ group is determined by elements $(g,\bar g,{\cT})$ with
multiplication
\begin{equation}
  \label{eq74} (g_1,\bar{g}_1,{\cT}_1)\cdot(g_2,\bar
g_2,{\cT}_2)=\big(g_1\circ g_2,\bar g_1\circ \bar
g_2,{\cT}_1+(g_1,\bar g_1) \cdot {\cT}_2\big),
\end{equation}
where the representation of $G$ on a weighted scalar $\eta^{s,w}$ is defined
through
\begin{equation}
  \big((g,\bar g)\cdot \eta^{s,w}\big)(x') = e^{w
E_R(x')} e^{-i s E_I (x')} \eta^{s,w}
(\xi,\bar{\xi}) .
\end{equation}
In the notation of section \ref{sec:coadj-repr-semi},
we now have $f=(g,\bar g)$, $X=(\cY, \bar{\cY})$, and $\alpha=\cT$
with
\begin{equation}
  \label{eq71a} \big(\sigma_f(\alpha)\big)(x')
  =\big((g,\bar g)\cdot {\cT}\big)(x')= e^{E_R(x')} {\cT}(x) .
\end{equation}
For the adjoint representation, we get
\begin{equation}
  \label{eq77} \big({\rm Ad}_f X\big)(x')=\big((g,\bar g)\cdot (\cY,\bar\cY)\big)
  (x')= \left(
e^{E_R(x')} e^{i E_I(x')} {\cY}(x),
e^{E_R(x')} e^{-i E_I(x')} {\bar{\cY}}(x) \right),
\end{equation}
\begin{equation}
  \label{eq78} \big(\Sigma_{{\rm Ad}_f X}\alpha\big)(x')= e^{E_R(x')}
({\mathcal{Y}}\mathcal{D} {\mathcal{T}} -
\frac{1}{2} \mathcal{D}
{\mathcal{Y}} {\mathcal{T}}  + {\rm c.c.})(x),
\end{equation}
whereas for the coadjoint representation, we get
\begin{equation} \big(Ad^*_{f}{\mathcal{J}}\big)(x') =
e^{-3E_R(x')} e^{iE_I(x')}{\mathcal{J}}(x) ,
\label{coadjoint group 1}
\end{equation}
\begin{equation}
  \big(\sigma^*_{f}{\mathcal{P}}\big)(x') =
e^{-3E_R(x')} {\mathcal{P}}(x) ,
\label{coadjoint group 2}
\end{equation}
\begin{equation}
  \big({\mathcal{T}} \times
\sigma^*_{f}{\mathcal{P}}\big)(x') = \left(
  e^{-3E_R(x')} e^{iE_I(x')} (\frac{1}{2}{\mathcal{T}} \bar{\mathcal{D}} {\mathcal{P}}  +\frac{3}{2} \bar{\mathcal{D}}
  {\mathcal{T}}{\mathcal{P}})(x)
, {\rm c.c.} \right) .
\label{coadjoint group 3}
\end{equation}

\subsection{Weyl invariance}
\label{sec:compl-weyl-resc}

The structure described above is covariant with respect to conformal coordinate
transformations combined with complex Weyl rescalings since it is defined in
terms of suitable covariant derivatives.

In particular, the above descriptions are valid for all conformal factors $P$
and $\bar P$ and all two-dimensional surfaces $\mathcal S$ such that total
$\mathcal{D},\bar{\mathcal{D}}$ derivatives are annihilated and the pairing is
non-degenerate. In the remainder of the paper, we mainly focus on two particular
cases: $(i)$ the sphere $S^2$ of radius $R$ with $P_S =
\frac{1+\xi\bar\xi}{R\sqrt 2} = \bar P_S$, and $(ii)$ the punctured complex
plane $\mathbb{C} - \{ 0 \} = \mathbb{C}_0$ with $P = 1 = \bar P$.

\section{Realization on the sphere}
\label{sec:Concrete realization on the sphere}

\subsection{Generalities}

If $\xi=\cot\frac{\theta}{2}e^{-i\phi}$, the standard metric on the sphere of
radius $R$ is
\begin{equation}
  ds^2=-2(P_S\bar P_S)^{-1}d\xi d\bar\xi,\quad
P_S=\frac{1+\xi\bar\xi}{R\sqrt 2} .
\label{eq:114}
\end{equation}
The globally well-defined conformal coordinate
transformations for the sphere are the fractional linear unimodular
transformations
\begin{equation}
  \label{eq:111} \xi'=\frac{a\xi +b}{c\xi+d},\quad ad-bc=1,\quad
a,b,c,d\in \mathbb C.
\end{equation}
In particular,
\begin{equation}
  \label{eq:30}
  \frac{\partial \xi}{\partial \xi'}=(c\xi+d)^2. 
\end{equation}
Under combined conformal coordinate transformations and Weyl rescalings, the
metric takes the standard form in the new coordinates if one freezes the Weyl
transformations as in equation \eqref{frozen factor expressions}. For $P_F=P_S$,
we have \cite{Newman1966,Held1970}
\begin{equation}
  \label{eq:112} e^{E(x')}=\frac{P_S(x)}{P_S(x')}\ddl{\xi'}{\xi} \iff
e^{E_R(x')}=\frac{1+\xi\bar\xi}{|a\xi+b|^2+|c\xi+d|^2},\quad e^{i
E_I(x')}=\frac{\bar c\bar\xi+\bar d}{c\xi+d}.
\end{equation}
In this context, $w$ is referred to as the boost weight.

The derivative operators \eqref{eq:34a} now take the explicit form (cf.~section
4.15 of \cite{Penrose:1984})
\begin{equation}
 \eth \eta^{s,w} = P_S^{1-s} \partial ( P_S^s
    \eta^{s,w} ), \quad \bar{\eth} \eta^{s,w} = P_S^{1+s} \bar\partial (P_S^{-s}
    \eta^{s,w}) .
\label{eth operators sphere}
\end{equation}
The pairing on the sphere is defined between scalars $\eta^{s,w}$ of weights $[s,w]$
and $\kappa^{s,-w-2}$ of weights $[s,-w-2]$ as follows,
\begin{equation}
  \label{eq60} \langle\kappa^{s,-w-2},\eta^{s,w}\rangle =\frac{1}{4\pi R^2}\int_{S^2} \frac{i
    d\xi\wedge d\bar{\xi}}{P_S \bar{P}_S}\ \widebar{\kappa^{s,-w-2}} \eta^{s,w}  ,
\end{equation}
where the normalization $C=(4\pi R^2)^{-1}$ is chosen so that
\begin{equation}
  \frac{1}{4\pi R^2}\int_{S^2} \frac{i
    d\xi\wedge d\bar{\xi}}{P_S \bar{P}_S}=\frac{1}{2\pi}\int_{S^2} \frac{i
    d\xi\wedge d\bar{\xi}}{(1+\xi\bar\xi)^2}
  = \frac{1}{4\pi}\int_0^\pi d\theta \sin\theta\int^{2\pi}_0 d\phi=1.
\label{eq:40}
\end{equation}
When compared to \eqref{eq:312}, we thus
have
\begin{equation}
  \label{eq61}  \langle([\cJ],[\bar\cJ],\cP),(\cY,\bar\cY,\cT)\rangle=\langle
  \cJ,\cY\rangle +\langle \bar\cJ,\bar\cY\rangle +\langle
  \cP,\cT\rangle,\quad d\mu(\xi,\bar\xi)=\frac{i d\xi
    \wedge d\bar{\xi}}{4\pi R^2 P_S \bar{P}_S}.
\end{equation}
This pairing has all the required properties.

\subsection{Adjoint and coadjoint representations of the group}
\label{sec:coadj-repr-group}

In terms of spin-weighted scalars, the (co)adjoint representation of the BMS$_4$
group on the sphere is described by the general formulas established in
subsection \ref{sec:Weighted scalars}, where now $f=(g,\bar g)$ are given by
general linear fractional transformations of \eqref{eq:111} (and the associated
transformation of the complex conjugate variable) and the factors $e^{E_R(x')}$
and $e^{iE_I(x')}$ are given in \eqref{eq:112}. Let us write them out
explicitly, with $x=\xi,\bar\xi$. 

For the adjoint representation, under a combined transformation $f$
and a supertranslation $\alpha$ with which one acts and a supertranslation
$\beta$ on which one acts, (where $\alpha,\beta$ are two different
supertranslation fields with the same weights than $\cT$ ),
\begin{equation}
  \label{eq:64}
  \begin{split}
    & \cY'(x')= e^{E_R(x')}e^{iE_I(x')}\cY(x),\\
    & \bar \cY'(x')= e^{E_R(x')}e^{-iE_I(x')}\bar\cY(x),\\
  & \beta'(x')=e^{E_R(x')}\Big(\beta-\big({\mathcal{Y}}\eth\alpha-  \frac{1}{2} \alpha 
\eth \mathcal{Y} + {\rm c.c.}\big)\Big)(x). 
\end{split}
\end{equation}

For the coadjoint representation, if we denote by $\cT$ instead of $\alpha$ the
supertranslation with which one acts, 
\begin{equation}
  \label{eq:65}
  \begin{split}
    &\cJ'(x')=e^{-3E_R(x')}e^{iE_I(x')}\Big(\cJ+ (\frac{1}{2}{\mathcal{T}}\bar\eth {\mathcal{P}}
    +\frac{3}{2}\bar\eth {\mathcal{T}}{\mathcal{P}}  )\Big)(x)\\
    &\bar \cJ'(x')=e^{-3E_R(x')}e^{-iE_I(x')}\Big(\bar \cJ+ (\frac{1}{2}{\mathcal{T}}\eth {\mathcal{P}} +
    \frac{3}{2}\eth
    {\mathcal{T}}{\mathcal{P}}  )\Big)(x)\\
    &\cP'(x')=e^{-3E_R(x')}\cP(x).
  \end{split}
\end{equation}

Not surprisingly, when using the associated conformal fields, these
transformations simplify. The formulas of section \ref{sec:Conformal fields}
apply. The Jacobians $\partial\xi/\partial\xi',\partial\bar\xi/\partial\bar\xi'$
are explicitly given by \eqref{eq:30} and its complex conjugate. In this case,
the integration measure is
\begin{equation}
  \tilde{d\mu}=\frac{id\xi\wedge d\bar\xi}{4\pi R^2},\label{eq:31}
\end{equation}
and
\begin{equation}
  \label{eq:64b}
  \begin{split}
    & \tilde\cY'(\xi')=(c\xi+d)^{-2}\tilde\cY(\xi),\\
    & \tilde{\bar\cY}'(\bar\xi')= (\bar c\bar\xi+\bar d)^{-2}\tilde{\bar\cY}(\xi),\\
    & \tilde \beta'(x')=(c\xi+d)^{-1}(\bar c\bar\xi+\bar d)^{-1}\Big(\tilde
    \beta-\big(\tilde{\mathcal{Y}}\partial\tilde \alpha- \frac{1}{2}
    \tilde\alpha \partial \tilde{\mathcal{Y}} + {\rm c.c.}\big)\Big)(x).
  \end{split}
\end{equation}
\begin{equation}
  \label{eq:65b}
  \begin{split}
    &\tilde \cJ'(x')=(c\xi+d)^{2}(\bar c\bar\xi+\bar d)^{4}\Big(\tilde\cJ(x)
    + (\frac{1}{2}\tilde{\mathcal{T}} \bar\partial \tilde{\mathcal{P}} +\frac{3}{2}\bar\partial
    \tilde{\mathcal{T}}\tilde{\mathcal{P}}  )\Big)(x)\\
    &\bar \cJ'(x')=(c\xi+d)^{4}(\bar c\bar\xi+\bar d)^{2}\Big(\tilde{\bar\cJ}
    + (\frac{1}{2}\tilde{\mathcal{T}} \partial \tilde{\mathcal{P}} +\frac{3}{2}\partial
    \tilde{\mathcal{T}} \tilde{\mathcal{P}}  )\Big)(x)\\
    &\tilde\cP'(x')=(c\xi+d)^{3}(\bar c\bar\xi+\bar d)^{3}\tilde\cP(x).
  \end{split}
\end{equation}

\subsection{Expansions}
\label{sec:expansions}

\subsubsection{Spin-weighted spherical harmonics}
\label{sec:spin-weight-spher}

We now decompose the relevant spin-weighted scalars in terms of spin-weighted
spherical harmonics (see appendix \ref{SWSH} for conventions).

For the $\mathfrak{bms}_4$ Lie algebra, we have $\bar{\mathcal{D}}\cY =
\bar\eth\cY=0$ with $\cY$ of weights $[-1,1]$. It follows from array (4.15.60)
of \cite{Penrose:1984} that $\cY$ belongs to the three-dimensional vector space
of spherical harmonics with spin weight $s=-1$ and $j=1$. Hence, defining
\begin{equation}
  \mathcal{Y}_m =   - R \sqrt{2}~ {}_{-1}Z_{1,m},\quad m=-1,0, 1,
  \label{eq:21}
\end{equation}
gives rise to the decomposition
\begin{equation}
  \bar\eth \cY=0\iff \cY = \sum_{m=-1}^{1} y_m \cY_m .
\label{Y decomposition}
\end{equation}

In the same way, 
\begin{equation}
  \bar{\cY}_m =  (-1)^m  R\sqrt{2} ~{}_{1}Z_{1,-m}\Longrightarrow  \bar \cY =
  \sum_{m=-1}^{1} \bar y_m \bar \cY_m, \label{eq:22}
\end{equation}
while
\begin{equation}
  \mathcal{T}_{j,m} = {}_{0}Z_{j,m} \Longrightarrow  \cT = \sum_{j, |m| \le j} t_{j,m} \mathcal{T}_{j,m}
  \label{T decomposition} ,
\end{equation}
where we impose $\bar t_{j,m} = (-1)^m t_{j,-m}$ since
$\cT$ is real.

When taking into account the pairing \eqref{eq60} and the normalization of the
${}_s Z_{jm}$ in \eqref{normZ}, this choice of basis for the algebra implies the
following choice for the associated dual basis of the coadjoint representation,
\begin{equation}
  \begin{split}
    \cY_*^m &= \frac{ - 6}{R\sqrt{2}(1+m)!(1-m)!}\ 
  {}_{-1}Z_{1,m},\label{eq:20}  \\
  \bar{\cY}_*^m & = \frac{ (-1)^m 6}{R\sqrt{2}(1+m)!(1-m)!}
  \ {}_{1}Z_{1,-m},\\
\mathcal{T}^{j,m}_* & =
\frac{(2j+1)!(2j)!}{j!j!(j+m)!(j-m)!}\ {}_{0}Z_{j,m},
  \end{split}
\end{equation}
and thus also the following expansions, 
\begin{equation}
  \cJ = \sum_{m=-1}^{1} j_m \cY_*^m , \quad \bar{\cJ} =
  \sum_{m=-1}^{1} \bar{j}_m \bar \cY_*^m ,\quad  \cP = \sum_{j, |m| \le j} p_{j,m}
  \mathcal{T}_*^{j,m}, 
\end{equation} 
where $\bar p_{j,m} = (-1)^m p_{j,-m}$ since $\cP$ is
real.

Note that the explicit expressions for ${}_{-1}Z_{1,m}$ in \eqref{def Z} gives
\begin{equation}
  \label{eq:90a}
  \tilde{\mathcal{Y}}_m=\mathcal{Y}_m P_S=  \xi^{1-m}, \quad
  \tilde{\bar\cY}_m  ={\bar\cY}_m \bar{P}_S=  {\bar\xi}^{1-m}.
\end{equation}

{\bf Remarks:}

(i) From the discussion of the behavior of spin-weighted spherical harmonics
under Lorentz transformations in section 4.15 of \cite{Penrose:1984}, it follows
that, if $w\geq |s|$ then $\eth^{w-s+1}\eta^{s,w}$, $\bar\eth^{w+s+1}\eta^{s,w} $ have definite spin and boosts
weights given by $[w+1,s-1]$ and $[-w-1,-s-1]$ respectively. This is the case
for
\begin{equation}
  \label{eq:124}
  \eth
  \bar\cY,\ \bar\eth\cY,\ \eth^3\cY,\ \bar\eth^3\bar\cY. 
\end{equation}
As shown there, the equations $\bar\eth\cY=0$ and $\eth^3\cY=0$ on the one hand,
and $\eth\bar\cY=0$ and $\bar\eth^3\bar\cY=0$ on the other, define the same
Lorentz invariant three-dimensional subspaces described above.

For the dual situation where $w\leq -|s|-2$, $\eth^{s-w-1}\kappa^{w+1,s-1}$ and
$\bar\eth^{-s-w-1}\kappa^{-w-1,-s-1}$ have definite spin and boost weights
$[s,w]$, it is shown that equivalence classes $[\eta^{s,w}], \eta^{s,w}\sim
\eta^{s,w}+\eth^{s-w-1}\kappa^{w+1,s-1}$ or $\eta^{s,w}\sim
\eta^{s,w}+\bar\eth^{-s-w-1}\kappa^{-w-1,-s-1}$ define Lorentz invariant
subspaces. This is the case for
\begin{equation}
  \label{eq:125}
  \bar\cJ\sim\bar\cJ+\bar \eth \bar \cL,\quad \bar\cJ\sim\bar\cJ+\eth^3\cM,
\end{equation}
where $\bar\cL:[2,-2]$ and $\cM:[-2,0]$ and both equivalence classes define the same
three-dimensional Lorentz invariant subspaces. Similarly, by complex conjugation
\begin{equation}
  \label{eq:126}
  \cJ\sim\cJ+\eth \cL,\quad\cJ\sim\cJ+\bar\eth^3\bar\cM
\end{equation}
where $\bar\cL:[-2,-2]$ and $\bar\cM:[2,0]$. 

(ii) The (well-known) coadjoint representation of the Poincar\'e group may be
discussed from the perspective developed here by imposing in addition the
conditions $\eth^2\cT=0=\bar\eth^2\cT$ reducing super to ordinary translations.
Again, these equations define a four-dimensional Lorentz invariant subspace
because $\cT$ has the required weights. At the same time, one should consider
equivalence classes $\cP\sim\cP +\eth^2 \cN+\xbar\eth^2 \xbar\cN$, where
$\cN:[-2,-1]$, $\bar\cN:[2,-1]$ have the required weights and which also defines a
Lorentz invariant four-dimensional subspace.

\subsubsection{Overcomplete set of functions}
\label{sec:overc-set-spin}

The representation of the generators $\mathcal Y_m,\bar{\mathcal Y}_m$ of the
Lorentz algebra on weighted scalars $\eta^{s,w}$ is explicitly given by
\begin{equation}
  \label{eq:89}
  \begin{split}
    \cY_m\cdot \eta^{s,w}&=   \xi^{-m}\big(\xi\partial \eta^{s,w}
    +(\frac{s-w}{2}(1-m)+w\frac{\xi\bar\xi}{1+\xi\bar\xi})\eta^{s,w}\big),\\
    \bar \cY_m\cdot \eta^{s,w}&=\bar\xi^{-m}\big(\bar\xi\bar\partial \eta^{s,w}
    +(-\frac{s+w}{2}(1-m)+w\frac{\xi\bar\xi}{1+\xi\bar\xi})\eta^{s,w}\big).
  \end{split}
\end{equation}
This follows from using \eqref{eq:49} written in terms of $\eth$ and $\bar\eth$
together with \eqref{eq:90a}.

For the associated conformal field $\phi_{h,\bar h}=P_S^w\eta^{s,w}$, this
simplifies to
\begin{equation}
  \label{eq:91}
  \begin{split}
    \tilde \cY_m\cdot \phi_{h,\bar h}&=\xi^{-m}\big(\xi\partial\phi_{h,\bar h}
    +h(1-m)\phi_{h,\bar h}\big),\\
    \tilde{\bar\cY}_m\cdot \phi_{h,\bar h}&=\bar\xi^{-m}\big(\bar\xi\bar\partial \phi_{h,\bar h}
    +\bar h(1-m)\phi_{h,\bar h}\big),
  \end{split}
\end{equation}
where $s=h-\bar h,w=-h-\bar h$.

Rather than expanding the spin-weighted scalar $\eta^{s,w}$ in terms of
(unnormalized) spin-weighted spherical harmonics, one may also work with
suitable sets of over-complete functions. We follow \cite{Goldberg1967}, section
4.C (up to conventions). Let $|s|\leq L$. For a fixed $L\in \mathbb N$, there is
an invertible matrix that relates the spin-weighted spherical harmonics
${}_sY_{j,m}$ with $j\leq L$ to the functions
\begin{equation}
{}_sZ^L_{m_1,m_2}=(1+\xi\bar\xi)^{-L}\xi^{L-s-m_1}\bar\xi^{L+s-m_2},\quad 0\leq
m_1\leq L-s,\quad 0\leq m_2\leq L+s \label{eq:95}. 
\end{equation}

Depending on the conformal weight $w$, one may label these same functions as
\begin{equation}
  \label{eq:98}
  {}_{h,\bar h} Z^{\tilde L}_{k,l}
  =P_S^{-w}{}_{h,\bar h} \tilde Z^{\tilde L}_{k,l},\quad   {}_{h,\bar h} \tilde Z^{\tilde L}_{k,l}
  = (R\sqrt 2)^{h+\bar h}(1+\xi\bar\xi)^{-\tilde L}\xi^{\tilde L-h-k}\bar\xi^{\tilde L-\bar h-l},
\end{equation}
where
\begin{equation}
  \label{eq:99}
  \tilde L=L+h+\bar h,\quad k=m_1+h,\quad l=m_2+\bar h,\quad h\leq k\leq \tilde L-h,\quad \bar h\leq l\leq \tilde L-\bar h. 
\end{equation}
In particular, if $h,\bar h$ are half-integer, so are $k,l$.

When taking for $\eta^{s,w}$ one of the functions ${}_{s,w}Z^{\tilde L}_{k,l}$, it follows
that
\begin{equation}
  \label{eq:97}
  \begin{split}
    \cY_m\cdot {}_{h,\bar h} Z^{\tilde L}_{k,l}&= -(hm+k){}_{h,\bar h}Z^{\tilde L+1}_{k+m,l}
     + (\tilde L-(hm+k)){}_{h,\bar h}Z^{\tilde L+1}_{k+m+1,l+1},\\
    \bar \cY_m\cdot {}_{h,\bar h} Z^{\tilde L}_{k,l}&= - (\bar hm+l){}_{h,\bar h}Z^{\tilde L+1}_{k,l+m}
     + (\tilde L-(\bar hm+l)){}_{h,\bar h}Z^{\tilde L+1}_{k+1,l+m+1},
\end{split}
\end{equation}
where the following (elementary) relations have been used,
\begin{equation}
  \label{eq:96}
  {}_{h,\bar h}Z^{\tilde L}_{k,l}={}_{h,\bar h}Z^{\tilde L+1}_{k,l}+{}_{h,\bar h}Z^{\tilde L+1}_{k+1,l+1}.
\end{equation}
By construction, when taking for the conformal fields $\phi_{h,\bar h}$ the
functions ${}_{h,\bar h} \tilde Z^{\tilde L}_{k,l}$, the relations \eqref{eq:97}
hold with the substitutions $\cY_m\to\tilde\cY_m$,
$\bar\cY_m\to\tilde{\bar\cY}_m$, ${}_{h,\bar h} Z^{\tilde L}_{k,l}\to {}_{h,\bar h}
\tilde Z^{\tilde L}_{k,l}$.

When taking into account that
\begin{equation}
  \label{eq:72}
\xi=\cot\frac{\theta}{2}e^{-i\phi},\quad  \mu=\cos\theta,\quad \xi\bar\xi=\frac{1+\mu}{1-\mu}=\cot^2\frac{\theta}{2},\quad
  1+\xi\bar\xi=\frac{2}{1-\mu},
\end{equation}
it follows that
\begin{equation}
  \label{eq:73}
  \begin{split}
  \langle{}_s Z^L_{m'_1,m'_2},{}_sZ^L_{m_1,m_2} \rangle&=\delta^{m_1+m_2'}_{m}\delta^{m_1'+m_2}_m\frac{1}{2}\int^1_{-1}d\mu
  (\frac{1-\mu}{2})^{2L}(\frac{1+\mu}{1-\mu})^{2L-m}\\
  &=\delta^{m_1+m_2'}_{m}\delta^{m_1'+m_2}_m\frac{m!(2L-m)!}{(2L+1)!}\\
  &  =\delta^{m_1+m_2'}_{m}\delta^{m_1'+m_2}_m\beta(m+1,2L-m+1),
\end{split}
\end{equation}
where $0\leq m \leq 2L$.
Instead of reverting to angular variables for the integrals, they may also be
worked out directly in complex coordinates:
\begin{multline}
  \label{eq:58}
  \langle{}_s Z^L_{m'_1,m'_2},{}_sZ^L_{m_1,m_2} \rangle =\frac{i}{2\pi}\int d\xi\wedge d\bar\xi\
  (1+\xi\bar\xi)^{-2L-2}\xi^{2L-m_1-m'_2}\bar\xi^{2L-m'_1-m_2}\\
  = \frac{1}{2L+1}\frac{1}{2\pi i}\int d\xi\wedge d\bar\xi\ \partial\big(
  (1+\xi\bar\xi)^{-2L-1}\bar\xi^{2L-m'_1-m_2-1}\big)\xi^{2L-m_1-m'_2},
\end{multline}
where $0\leq m'_1+m_2\leq 2L$, $0\leq m_1+m'_2\leq 2L$. If $m_1+m'_2=2L$, one then
proceeds by using Stokes' theorem together with a kind of Cauchy residue theorem
for the remaining line integral (see for instance \cite{Porter:1981cg} in the
current context). If $m_1+m'_2<2L$, one makes an integration by parts to lower
the degree of $\xi$, and applies the same reasoning for all integrals that
involve a total $\partial$ with expressions that have poles in $\bar \xi$.

Because of the weights $[s,w]$ and $[s,-w-2]$ of the spin-weighted spherical
harmonics involved in \eqref{eq60}, the relevant integrals pair ${}_{h,\bar
  h}Z^{\tilde L}_{k,l}$ on the right with ${}_{h',\bar h'}Z^{\tilde L'}_{k',l'}$
on the left, where $h'=-\bar h+1$, $\bar h'=-h+1$ and $\tilde L'=\tilde
L-2h-2\bar h+2$,
\begin{equation}
  \label{eq:102}
  \langle {}_{-\bar h+1,-h+1}Z^{\tilde L-2 h-2 \bar h+2}_{k',l'},{}_{h,\bar h}Z^{\tilde L}_{k,l} \rangle
  =\delta^{k+l'-1}_m\delta^{k'+l-1}_m\beta(m+1,2(\tilde L-h-\bar h)-m+1).
\end{equation}

By construction, the associated conformal fields ${}_{h,\bar h}\tilde Z^{\tilde
  L}_{k,l}$ have the same integrals when using the measure $\tilde{d\mu}$ given in
\eqref{eq:31}, 
\begin{equation}
  \label{eq:44}
  \langle {}_{-\bar h+1,-h+1}\tilde Z^{\tilde L-2 h-2 \bar h+2}_{k',l'},{}_{h,\bar h}\tilde Z^{\tilde L}_{k,l} \tilde\rangle
  =\delta^{k+l'-1}_m\delta^{k'+l-1}_m\beta(m+1,2(\tilde L-h-\bar h)-m+1).
\end{equation}

\subsection{Structure constants}
\label{sec:bases-struct-const}

When taking as generators for supertranslations the unnormalized spherical
harmonics \eqref{T decomposition}, one can now work out the structure constants
of the $\mathfrak{bms}_4$ algebra \eqref{algebrabms}-\eqref{commutator0} by
using properties \eqref{eth acting on Z} and \eqref{product of SWSH}. For the
first part of the algebra, they can be read off from the commutation relations
\begin{equation}
 [\cY_m, \cY_n] = (m-n) \cY_{m+n}, \quad [\bar \cY_m,
\bar \cY_n] = (m-n) \bar \cY_{m+n}, \quad [\cY_m, \bar \cY_n] = 0,
\label{de witt part}
\end{equation}
while those involving supertranslation generators can be obtained
from
\begin{multline}
  [\mathcal{Y}_{-1} , \mathcal{T}_{j,m}] =
   +  \frac{(j+2)(j+m)(j+m-1)}{4(2j+1)(2j-1)} \mathcal{T}_{j-1,m-1}   -  
  \frac{(j+m)}{2} \mathcal{T}_{j,m-1} \\   +  (j-1)
  \mathcal{T}_{j+1,m-1}, \label{supertranslation 1}
\end{multline}
\begin{flalign}
  [\mathcal{Y}_{0}, \mathcal{T}_{j,m}] =-\frac{(j+2)(j+m)(j-m)}{4(2j+1)(2j-1)}
\mathcal{T}_{j-1,m} - \frac{m}{2} \mathcal{T}_{j,m} + (j-1)
\mathcal{T}_{j+1,m} ,
\label{supertranslation 2}
\end{flalign}
\begin{multline}
[\mathcal{Y}_{1} ,
\mathcal{T}_{j,m}] =   + \frac{(j+2)(j-m)(j-m-1)}{4(2j+1)(2j-1)}
\mathcal{T}_{j-1,m+1}   +    \frac{(j-m)}{2} \mathcal{T}_{j,m+1} \\
  +    (j-1) \mathcal{T}_{j+1,m+1} ,
\label{supertranslation 3}
\end{multline}
The commutation relations involving $\bar{\mathcal{Y}}_m$ and $\mathcal{T}_{j,m}$
may then be obtained by complex conjugation. They are explicitly given by 
\begin{multline}
[\bar{\mathcal{Y}}_{-1} ,
\mathcal{T}_{j,m}] =   -   \frac{(j+2)(j-m)(j-m-1)}{4(2j+1)(2j-1)}
\mathcal{T}_{j-1,m+1}  +   \frac{(j-m)}{2} \mathcal{T}_{j,m+1} \\  -    (j-1)
\mathcal{T}_{j+1,m+1} ,
\label{supertranslation 4}
\end{multline}
\begin{equation}
[\bar{\mathcal{Y}}_{0} ,
\mathcal{T}_{j,m}] = -\frac{(j+2)(j+m)(j-m)}{4(2j+1)(2j-1)}
\mathcal{T}_{j-1,m} + \frac{m}{2} \mathcal{T}_{j,m} + (j-1)
\mathcal{T}_{j+1,m} ,
\label{supertranslation 5}
\end{equation}
\begin{multline} [\bar{\mathcal{Y}}_{1} , \mathcal{T}_{j,m}] = -
  \frac{(j+2)(j+m)(j+m-1)}{4(2j+1)(2j-1)}
  \mathcal{T}_{j-1,m-1}   -   \frac{(j+m)}{2} \mathcal{T}_{j,m-1} \\
  - (j-1) \mathcal{T}_{j+1,m-1} .
\label{supertranslation 6}
\end{multline}
Finally, the supertranslations generators commute with each other,
\begin{equation}
[\mathcal{T}_{j,m} , \mathcal{T}_{j',m'}] =0\label{eq:23}.
\end{equation}

In order to establish the relation to the commutation relations of section
\ref{Relations with literature}, one defines
\begin{equation}
l_m = \tilde{\mathcal{Y}}_m \partial ,\quad \bar{l}_m 
= \tilde{{\bar\cY}}_m \bar \partial,
\end{equation}
and takes into account equation \eqref{eq:90a} together with $\xi=\cot
\frac{\theta}{2} e^{-i\phi}$. If one makes the identification of the generators
as in \eqref{eq:33} at $r\to \infty$ and $u=0$, it follows that
\begin{align}
   & L_z=-i(l_0-\bar l_0)=-i(\xi\d-\bar\xi\bar\d),
   &  K_z=-(l_0+\bar l_0)=-(\xi\d+\bar\xi\bar\d),\nonumber \\
   & L^+=  +  (l_{1}+\bar l_{-1})=\d+\bar\xi^2\bar \d,
   & L^-=  +   (\bar l_{1}+ l_{-1})=\bar \d+\xi^2\d,  \label{eq:28}\\
   & K^+=  - (\bar l_{-1}-l_1)=\d-\bar \xi^2\bar\d,
   & K^-=  -   (l_{-1}-\bar l_1)=\bar \d-\xi^2\d.\nonumber 
\end{align}
This allows one to explicitly relate the
commutation relations for the Lorentz algebra in \eqref{de witt part} to those
in \eqref{Lorentz algebra}, respectively to the first part of \eqref{eq:29}. The
Poincar\'e generators are represented by
\begin{equation}
  \label{eq:39}
  H=1,\quad P_z=\frac{1-\xi\bar\xi}{2(1+\xi\bar\xi)},\quad P^+
    =-\frac{\bar\xi}{{1+\xi\bar\xi}},\quad P^-=\frac{\xi}{{1+\xi\bar\xi}}.
\end{equation}
For functions $f$ on the sphere, we now get instead of \eqref{eq:34}
\begin{equation}
  \label{eq:38}
  \begin{split}
    [L_z,f]&=L_z(f),\quad [L^\pm,f]=L^\pm (f),\quad
    [K_z,f]=K_z(f)+\frac{1-\xi\bar\xi}{1+\xi\bar\xi} f,
    \\
    [K^+,f]& =K^+(f)+\frac{2\bar\xi}{1+\xi\bar\xi} f,\quad
    [K^-,f]=K^-(f)+\frac{2\xi}{1+\xi\bar\xi} f.
\end{split}
\end{equation}
When applied to the four Poincar\'e generators in \eqref{eq:39}, this
reproduces the second part of \eqref{eq:29}.

In order to relate the action of the Lorentz generators on the supertranslation
generators \eqref{T decomposition} given in \eqref{supertranslation
  1}-\eqref{supertranslation 6} to the more standard form
\eqref{commutators1}-\eqref{commutators6}, we may start from the (first equality
in the) relations \eqref{eq:28} and use \eqref{supertranslation
  1}-\eqref{supertranslation 6} to show that
\begin{equation}
  \begin{split}
    [L_z, \mathcal T_{j,m}] =& i m \mathcal T_{j,m},\quad [L^\pm, \mathcal
    T_{j,m}]
    = \pm (j\mp m) \mathcal T_{j,m\pm 1},\\
    [K_z, \mathcal T_{j,m}] =& -2(j-1) \mathcal T_{j+1,m} +
    \frac{(j+2)(j+m)(j-m)}{2(2j+1)(2j-1)} \mathcal
    T_{j-1,m} ,\\ [K^\pm, \mathcal
    T_{j,m}]
    =& \pm 2 (j-1) \mathcal T_{j+1,m\pm 1} \pm \frac{(j+2)(j\mp m)(j\mp m -1)}{2(2j+1)(2j-1)}
\mathcal T_{j-1,m\pm 1} .
\end{split}
\end{equation}
When taking the normalization \eqref{Y in term of Z} into account, we then
recover the commutation relations \eqref{commutators1}-\eqref{commutators6}.
Note that the commutation relations of $L_z$ with the supertranslations
generators are particularly simple since the latter are expressed in terms of
(unnormalized) spherical harmonics.

The choice of basis for the Lorentz algebra in \eqref{de witt part} is adapted
to the $\mathfrak{sl}(2,\mathbb{R})\times \mathfrak{sl}(2,\mathbb{R})$
decomposition. It is thus useful to organize the supertranslation generators, or
more generally, the functions on the sphere, accordingly. This will also allow
us to compare directly with the realization on the punctured complex plane to be
discussed below. Hence, instead of providing the commutation relation between
the Lorentz and supertranslation generators, one may replace the latter by the
overcomplete set of functions adapted to $\cT$ of weights $[0,1]$,
\begin{equation}
  \cT^{\tilde L}_{k,l}={}_{-\frac{1}{2},-\frac{1}{2}}Z^{\tilde L}_{k,l}\label{eq:100}.
\end{equation}
One then finds 
\begin{equation}
  \label{eq:94}
  \begin{split}
    [\cY_m,\cT^{\tilde L}_{k,l}]=(\frac{m}{2}-k)\cT^{\tilde L+1}_{k+m,l}
    +(\tilde L +\frac{m}{2}-k)\cT^{\tilde L+1}_{k+m+1,l+1},\\
    [\bar \cY_m,\cT^{\tilde L}_{k,l}]=(\frac{m}{2}-l)\cT^{\tilde L+1}_{k,l+m}
    +(\tilde L +\frac{m}{2}-l)\cT^{\tilde L+1}_{k+1,l+m+1},
  \end{split}
\end{equation}
\begin{equation}
  \label{eq:105}
  [\mathcal T^{\tilde L}_{k,l},\mathcal T^{\tilde L}_{k',l'}]=0. 
\end{equation}

\subsection{Coadjoint representation of the algebra}
\label{sec:coadj-repr-algebra}

The coadjoint representation may now be written explicitly using \eqref{eq:3}.
Alternatively, it can be derived using the results of subsection
\ref{sec:algebra-1} together with \eqref{eth acting on Z} and \eqref{product of
  SWSH}. One finds,
\begin{equation}
{\rm ad}^*_{\mathcal Y_m} \mathcal{Y}^n_* = (-2m+n)
\mathcal{Y}^{n-m}_* , \quad 
{\rm ad}^*_{\mathcal Y_m} \bar{\mathcal{Y}}^n_* = 0,
\end{equation}
\begin{equation}
  \label{eq:120}
  {\rm ad}^*_{\bar\cY_m}
  \bar{\mathcal{Y}}^n_* = (-2m+n) \bar{\mathcal{Y}}^{n-m}_* , \quad {\rm ad}^*_{\bar\cY_m}
  \mathcal{Y}^n_*=0,
\end{equation}
\begin{align} {\rm ad}^*_{\mathcal Y_{-1}} \mathcal{T}^{j,m}_* =
  & -  \frac{(j+3)(j+m+2)(j+m+1)}{4(2j+3)(2j+1)} \mathcal{T}^{j+1,m+1}_*
    \nonumber \\ &\qquad\qquad\qquad\qquad  +  \frac{(j+m+1)}{2}
                   \mathcal{T}^{j,m+1}_*  -   (j-2) \mathcal{T}^{j-1,m+1}_* ,\label{eq1} \\ &\nonumber
  \\{\rm ad}^*_{\mathcal Y_0} \mathcal{T}^{j,m}_* =
&\frac{(j+3)(j+m+1)(j-m+1)}{4(2j+3)(2j+1)} \mathcal{T}^{j+1,m}_*
+\frac{m}{2} \mathcal{T}^{j,m}_* -(j-2) \mathcal{T}^{j-1,m}_* , \\
  {\rm ad}^*_{\mathcal Y_1} \mathcal{T}^{j,m}_* =
  &  -   \frac{(j+3)(j-m+2)(j-m+1)}{4(2j+3)(2j+1)} \mathcal{T}^{j+1,m-1}_*
    \nonumber \\ &\qquad\qquad\qquad\qquad  -   \frac{(j-m+1)}{2}
                   \mathcal{T}^{j,m-1}_*   -   (j-2) \mathcal{T}^{j-1,m-1}_* , \\ &\nonumber
  \\ {\rm ad}^*_{\bar\cY_{-1}} \mathcal{T}^{j,m}_* = &  +   \frac{(j+3)(j-m+2)(j-m+1)}{4 (2j+3)(2j+1)} \mathcal{T}^{j+1,m-1}_*
    \nonumber \\ &\qquad\qquad\qquad\qquad   -    \frac{(j-m+1)}{2}
                   \mathcal{T}^{j,m-1}_*   +    (j-2) \mathcal{T}^{j-1,m-1}_* , \\ &\nonumber
  \\{\rm ad}^*_{\bar\cY_0} \mathcal{T}^{j,m}_* = &+
\frac{(j+3)(j+m+1)(j-m+1)}{4 (2j+3) (2j+1)} \mathcal{T}^{j+1,m}_* -
\frac{m}{2} \mathcal{T}^{j,m}_* - (j-2) \mathcal{T}^{j-1,m}_* , \\
  {\rm ad}^*_{\bar\cY_1} \mathcal{T}^{j,m}_* = &  +   \frac{(j+3)(j+m+2)
(j+m+1)}{4 (2j +3) (2j+1)} \mathcal{T}^{j+1,m+1}_* \nonumber \\
  &\qquad\qquad\qquad\qquad   +   \frac{(j+m+1)}{2} \mathcal{T}^{j,m+1}_*   + 
(j-2) \mathcal{T}^{j-1,m+1}_* ,\label{eq2}
\end{align}

\begin{equation} {\rm ad}^*_{\mathcal{T}_{j,m}} \mathcal{Y}^p_*  = 0 =
  {\rm ad}^*_{\mathcal{T}_{j,m}} \bar{\mathcal{Y}}^p_*  ,\label{trivial}
\end{equation}

\begin{equation}
  \label{eq:92}
  \begin{split}
    &{\rm ad}^*_{\mathcal{T}_{j,m}}
    \mathcal{T}_*^{j',m'}=\\
    & \big(-\frac{(j+2)(j+m)(j+m-1)}{4(2j+1)(2j-1)}\delta^{j'}_{j-1}
    +\frac{(j+m)}{2}\delta^{j'}_j-
    (j-1)\delta^{j'}_{j+1}\big)\delta^{m'}_{m-1}\mathcal
    Y_*^{-1}\\
    &+\big(-\frac{(j+2)(j+m)(j-m)}{4(2j+1)(2j-1)}\delta^{j'}_{j-1}-
    \frac{m}{2}\delta^{j'}_j + (j-1)\delta^{j'}_{j+1}\big)\delta^{m'}_m\mathcal
    Y_*^{0}\\
    &+\big(-\frac{(j+2)(j-m)(j-m-1)}{4(2j+1)(2j-1)}\delta^{j'}_{j-1} -
    \frac{(j-m)}{2}\delta^{j'}_j-
    (j-1)\delta^{j'}_{j+1}\big)\delta^{m'}_{m+1}\mathcal
    Y_*^{1}\\
    &+\big(+\frac{(j+2)(j-m)(j-m-1)}{4(2j+1)(2j-1)}\delta^{j'}_{j-1}
    - \frac{(j-m)}{2}\delta^{j'}_j+ (j-1)\delta^{j'}_{j+1}\big)\delta^{m'}_{m+1}\bar{\mathcal{Y}}^{-1}_*\\
    &+\big(-\frac{(j+2)(j+m)(j-m)}{4(2j+1)(2j-1)}\delta^{j'}_{j-1}+
    \frac{m}{2}\delta^{j'}_j+ (j-1)\delta^{j'}_{j+1}\big)
    \delta^{m'}_m\bar{\mathcal{Y}}^{0}_*\\
    &+\big(+\frac{(j+2)(j+m)(j+m-1)}{4(2j+1)(2j-1)}\delta^{j'}_{j-1}+\frac{(j+m)}{2}\delta^{j'}_j
    + (j-1)\delta^{j'}_{j+1}\big)\delta^{m'}_{m-1}\bar{\mathcal{Y}}^{1}_*.
\end{split}
\end{equation}

In terms of the overcomplete sets of functions, if one uses
${}_{\frac{3}{2},\frac{3}{2}}Z^{\tilde L+4}_{k,l}$ rather than $\mathcal
T_*^{j,m}$ for the expansion of $\mathcal P$, one may use \eqref{eq:97} to
replace equations \eqref{eq1}-\eqref{eq2} through
\begin{equation}
  \label{eq:97supermomenta}
  \begin{split}
    \cY_m\cdot {}_{\frac{3}{2},\frac{3}{2}}Z^{\tilde L+4}_{k,l}&= -
    (\frac{3}{2}m+k)\mathcal {}_{\frac{3}{2},\frac{3}{2}}Z^{\tilde L+5}_{k+m,l}
    +   (\tilde L+4-(\frac{3}{2}m+k)){}_{\frac{3}{2},\frac{3}{2}}Z^{\tilde L+5}_{k+m+1,l+1},\\
    \bar \cY_m\cdot {}_{\frac{3}{2},\frac{3}{2}}Z^{\tilde L+4}_{k,l}&= -
    (\frac{3}{2}m+l){}_{\frac{3}{2},\frac{3}{2}}Z^{\tilde L+5}_{k,l+m} + (\tilde
    L+4-(\frac{3}{2}m+l)){}_{\frac{3}{2},\frac{3}{2}}Z^{\tilde L+5}_{k+1,l+m+1},
  \end{split}
\end{equation}
while equations \eqref{trivial} become 
\begin{equation}
  \label{eq:106}
  {\rm ad}^*_{\mathcal T^{\tilde L}_{k,l}}  \mathcal Y^p_*=0={\rm ad}^*_{\mathcal T^{\tilde L}_{k,l}}  \bar{\mathcal Y}^p_*.
\end{equation}
Finally, it also follows from
\begin{equation}
  \label{eq:104}
  \langle {}_{\frac{3}{2},\frac{3}{2}}Z^{\tilde L+4}_{k',l'},\mathcal T^{\tilde L}_{k,l} \rangle
  =\delta^{k'+l-1}_m\delta^{k+l'-1}_m\beta(m+1,2\tilde L+3-m),
\end{equation}
the commutation relations \eqref{eq:94}, \eqref{eq:105} and the definition of
the coadjoint representation that
\begin{multline}
  \label{eq:107}
  {\rm ad}^*_{\mathcal T^{\tilde L}_{k,l}}\ {}_{\frac{3}{2},\frac{3}{2}}Z^{\tilde L+5}_{k',l'}
  =\big(\frac{k'-3k+l-l'}{2}\beta(k'+l,2\tilde L+6-k'-l)\\
  +(\tilde L+\frac{k'-3k+l-l'}{2})\beta(k'+l+1,2\tilde L+5-k'-l)\big)\cY_*^{k'-k+l-l'}\\
  +\big(\frac{l'-3l+k-k'}{2}\beta(l'+k,2\tilde L+6-l'-k)\\
  +(\tilde L+\frac{l'-3l+k-k'}{2})\beta(l'+k+1,2\tilde
  L+5-l'-k)\big)\bar{\cY}_*^{l'-l+k-k'}.
\end{multline}

\section{Realization on the punctured complex plane}
\label{sec:Concrete realization on the two-punctured Riemann sphere}

\subsection{Generalities}

Since the whole structure is Weyl invariant, one may start from the sphere with
radius $R$ and perform a Weyl rescaling as in \eqref{Pure Weyl} with
\begin{equation}
  \label{eq:67}
  e^{-E(\xi,\bar\xi)}=\frac{\sqrt 2}{1+\xi\bar\xi}, 
\end{equation}
followed by the (conformal) coordinate transformations that consists of a simple
rescaling $\xi=R^{-1}z$, $\bar \xi =R^{-1}\bar z$, so that the metric becomes 
\begin{equation}
  ds^2 = -2 d z d \bar{z}.
\end{equation}

The next step is to remove the points at infinity and at the origin to go to the
1-punctured complex plane $\mathbb{C}_0$. This changes the allowed space of
functions. Conformal coordinate transformations are of the form
\begin{equation}
   z'=z'(z), \quad \bar z'=\bar z'(\bar z),
 \end{equation}
 where the globally well-defined ones that are connected to the identity are
 $z'=az, a\in \mathbb C, a\neq 0$. The derivative operators $\eth$ and
 $\bar{\eth}$ defined in \eqref{eq:34a} simply become $\partial$ and
 $\bar{\partial}$, respectively. There is no difference between conformal fields
 and weighted scalars. Indeed, freezing the conformal factor as in \eqref{frozen
   factor expressions} with $P_F = 1 = \bar{P}_F$ yields
\begin{equation}
  e^{E(x')} = \frac{\partial z'}{\partial z} ~
\Longleftrightarrow ~ e^{E_R(x')} = \left( \frac{\partial
z'}{\partial z} \frac{\partial \bar{z}'}{\partial \bar{z}}
\right)^{\frac{1}{2}},~ e^{iE_I(x')} = \left(\frac{\partial z' /
\partial z }{\partial \bar{z}' / \partial \bar{z}}
\right)^{\frac{1}{2}} ,
\label{fixed P=1}
\end{equation}
which implies that conformal fields (of vanishing Weyl weights) and their
associated weighted scalars through the map \eqref{mapping} are equal and
transform in the same way (compare \eqref{eq:93} and \eqref{eq:109} by taking
\eqref{fixed P=1} into account). In the following, we use the notation for
conformal fields $\phi_{h, \bar{h}}$.

We assume here that conformal fields on the punctured complex plane may be
expanded in series as
\begin{equation}
   \phi_{h,\bar{h}} (z, \bar{z}) = \sum_{k, l } a_{k,l}\ {}_{h,\bar h} \tilde Z_{k,l},\quad {}_{h,\bar h} \tilde
  Z_{k,l}=z^{-h-k} \bar{z}^{-\bar{h}-l},
\label{Laurent series}
\end{equation}
where the coefficients $a_{k,l} \in \mathbb{C}$ and satisfy suitable conditions
that we will not discuss in detail here (see e.g.~\cite{Kac1996} for more
details). We also assume that $h,\bar h$ are either integer or half-integer. In
the former case $k,l\in \mathbb Z$, whereas in the latter case $k,l\in
\frac{1}{2} +\mathbb Z$. Other choices are also possible. The reason we are
choosing Neveu-Schwarz conditions here is that, up to factors of $(1+z\bar z)$,
the functions that appear here then include those that have appeared naturally
in the case of the sphere.

Residues with respect to $z$ and $\bar{z}$ are defined as
\begin{equation} 
\text{Res}_{z} [ \phi_{h,\bar{h}}] (\bar{z}) =
\sum_{l} a_{1-h,l}\bar{z}^{-\bar h-l}, \quad
\text{Res}_{\bar{z}} [ \phi_{h,\bar{h}}] (z) = \sum_{k} a_{k,1 - \bar{h}} z^{-h-k}.
\end{equation}
This allows one to define pairing
\begin{equation}
  \langle  \psi_{-\bar h+1,-h+1},  \phi_{h,\bar h} \tilde \rangle = \text{Res}_{z}
\text{Res}_{\bar{z}} [\overline{ \psi_{-\bar h+1,-h+1}}  \phi_{h,\bar h}] .
\end{equation}
This pairing is non-degenerate, and since $\text{Res}_{z} [\partial
\phi] =0 = \text{Res}_{\bar{z}} [\bar{\partial}\phi]$, it annihilates total
derivatives $\partial$ and $\bar{\partial}$, as it should.
The pairing can then be defined as
\begin{equation}
  \langle([\tilde{\cJ}],[\tilde{\bar\cJ}],\tilde{\cP}),
  (\tilde{\cY},\tilde{\bar\cY},\tilde{\cT})\tilde\rangle=\langle
  \tilde{\cJ},\tilde{\cY}\tilde\rangle +\langle
  \tilde{\bar\cJ},\tilde{\bar\cY}\tilde\rangle +\langle
  \tilde{\cP},\tilde{\cT}\tilde\rangle .
\end{equation}

\subsection{Adjoint and coadjoint representations of the group}
\label{sec:adjo-coadj-repr}

The formulas for the adjoint and coadjoint representations of the group are the
same than those for the conformal fields on the sphere, except for the general
Jacobians $\partial z/\partial z',\partial\bar z/\partial\bar z'$,
\begin{equation}
  \label{eq:64c}
  \begin{split}
    & \tilde\cY'(z')=\big(\frac{\partial z}{\partial z'}\big)^{-1}\tilde\cY(z),\\
    & \tilde{\bar\cY}'(\bar z')= \big(\frac{\partial \bar z}{\partial \bar z'}\big)^{-1}\tilde{\bar\cY}(z),\\
    & \tilde \beta'(x')=\big(\frac{\partial z}{\partial z'}\big)^{-\frac{1}{2}}
    \big(\frac{\partial \bar z}{\partial \bar z'}\big)^{-\frac{1}{2}}\Big(\tilde
    \beta-\big(\tilde{\mathcal{Y}}\partial\tilde \alpha- \frac{1}{2}
    \tilde\alpha \partial \tilde{\mathcal{Y}} + {\rm c.c.}\big)\Big)(x),
  \end{split}
\end{equation}
\begin{equation}
  \label{eq:65c}
  \begin{split}
    &\tilde \cJ'(x')=\big(\frac{\partial z}{\partial z'}\big)^{1}
    \big(\frac{\partial \bar z}{\partial \bar z'}\big)^{2}\Big(\tilde\cJ+
    (\frac{1}{2}\tilde{\mathcal{T}} \bar\partial \tilde{\mathcal{P}} +\frac{3}{2}
    \bar\partial
    \tilde{\mathcal{T}}
    \tilde{\mathcal{P}}  )\Big)(x)\\
    &\tilde{\bar\cJ}'(x')=\big(\frac{\partial z}{\partial z'}\big)^{2}
    \big(\frac{\partial \bar z}{\partial \bar z'}\big)^{1}\Big(\tilde{\bar\cJ}+
    (\frac{1}{2} \tilde{\mathcal{T}} \partial \tilde{\mathcal{P}} +
    \frac{3}{2}\partial
    \tilde{\mathcal{T}} \tilde{\mathcal{P}}  )\Big)(x)\\
    &\tilde\cP'(x')=\big(\frac{\partial z}{\partial z'}\big)^{\frac{3}{2}}
    \big(\frac{\partial \bar z}{\partial \bar
      z'}\big)^{\frac{3}{2}}\tilde\cP(x).
  \end{split}
\end{equation}

\subsection{Expansions}
\label{sec:expansions-1}

In terms of the basis functions defined in \eqref{Laurent series},  
\begin{equation}
  \label{eq:60}
  \langle{}_{-\bar h+1,-h+1} \tilde Z_{k',l'}, {}_{h,\bar h} \tilde Z_{k,l} \tilde\rangle=\delta_{l'+k}^0\delta_{k'+l}^0. 
\end{equation}
In particular, the dual becomes 
\begin{equation}
  ({}_{h,\bar h} \tilde Z_{k,l})^*={}_{-\bar h+1,-h+1} \tilde Z_{-l,-k}\iff (z^{-h-k}\bar z^{-\bar h-l})^*
  =z^{\bar h -1+l}\bar z^{h-1+k}\label{eq:61}. 
\end{equation}

The basis for the conformal fields relevant for the algebra is
\begin{equation}
  \tilde{\cY}_m= {}_{-1,0}\tilde Z_{m,0}=z^{1-m} , \quad  \tilde{\bar\cY}_m ={}_{0,-1}\tilde Z_{0,m}= 
  \bar{z}^{1-m} , \quad \mathscr{\tilde T}_{k,l} = {}_{-\frac{1}{2},-\frac{1}{2}}
  \tilde Z_{k,l}=z^{\frac{1}{2}-k}
  \bar{z}^{\frac{1}{2}-l}, \label{conformal basis riemann}
\end{equation}
where $m,k+\frac{1}{2},l+\frac{1}{2} \in \mathbb{Z}$. We have
\begin{equation} 
  \tilde{\cY}= \sum_{m \in \mathbb{Z}} \tilde{y}_m
  \tilde{\cY}_m , \quad \tilde{\bar{\cY}}= \sum_{m \in \mathbb{Z}}
  \tilde{\bar{y}}_m \tilde{\bar\cY}_m , \quad \tilde{\mathcal{T}} = \sum_{k,l
\in \frac{1}{2}+\mathbb{Z}} \tilde{t}_{k,l} \mathscr{\tilde T}_{k,l} , \label{decomp gen
riemann}
\end{equation}
where $\bar{\tilde{t}}_{k,l} = \tilde{t}_{l,k}$ since $\tilde{\mathcal{T}}$ is
real. For the coadjoint representation, one finds from \eqref{eq:61} (or from
the definition with equivalence classes when taking into account that
$z^{-1},\bar z^{-1}$ are not equivalent to zero because they are not the
derivative of a monomial but of the logarithm),
\begin{equation}
\tilde{\cY}^m_* =
z^{-1} {\bar{z}}^{-2+m},\quad\tilde{\bar\cY}_*^m = {z}^{-2+m}
\bar{z}^{-1},\quad\mathscr{\tilde T}^{k,l}_* = {z}^{-\frac{3}{2}+l}
{\bar{z}}^{-\frac{3}{2}+k},\label{eq:59}
\end{equation}
we have
\begin{equation}
  \tilde{\cJ}= \sum_{m \in \mathbb{Z}} \tilde{j}_m
  \tilde{\cY}^m_* , \quad \tilde{\bar{\cJ}}= \sum_{m \in \mathbb{Z}}
  \tilde{\bar{j}}_m \tilde{\bar\cY}_*^m , \quad \tilde{\mathcal{P}} =
\sum_{k,l \in \frac{1}{2}+\mathbb{Z}} \tilde{p}_{k,l} \mathscr{\tilde T}^{k,l}_*
, \label{decomp gen riemann 2}
\end{equation}
where $\bar{\tilde{p}}_{k,l} = p_{l,k}$ since
$\tilde{\mathcal{P}}$ is real.

In terms of basis elements, the representation \eqref{eq:49a} becomes
\begin{equation}
  \label{eq:62}
  \tilde\cY_m\cdot {}_{h,\bar h} \tilde Z_{k,l}=-(hm+k)\ {}_{h,\bar h} \tilde Z_{k+m,l},\quad
  \tilde{\bar\cY}_m\cdot {}_{h,\bar h} \tilde Z_{k,l}=-(\bar hm+l)\ {}_{h,\bar h} \tilde Z_{k,l+m},
\end{equation}
while
\begin{equation}
  \label{eq:63}
  \begin{split}
  \tilde\cY_m\cdot ({}_{h,\bar h} \tilde Z_{k,l})^*&=[(\bar h-1)m+l]\ ({}_{h,\bar h} \tilde Z_{k,l-m})^*,\\
  \tilde{\bar\cY}_m\cdot ({}_{h,\bar h} \tilde Z_{k,l})^*&=[(h-1)m+k]\ ({}_{h,\bar h}
  \tilde Z_{k-m,l})^*.
\end{split}
\end{equation}

\subsection{Structure constants}

As discussed in subsection \ref{sec:algebra-1}, all the results stated
there can be readily expressed in terms of conformal fields. In
particular, the $\mathfrak{bms}_4$ algebra 
\eqref{algebrabms}-\eqref{commutator0} simplifies to
\begin{equation}
  [(\tilde{\cY}_1, \tilde{\bar{\cY}}_1, \tilde{\cT}_1),
(\tilde{\cY}_2, \tilde{\bar{\cY}}_2, \tilde{\cT}_2) ] =
(\hat{\tilde{\cY}}, \hat{\tilde{\bar{\cY}}}, \hat{\tilde{\cT}}),
\end{equation}
where
\begin{equation} \left\{
      \begin{aligned} \hat{\tilde{\cY}} &= \tilde{\cY}_1 \partial
\tilde{\cY}_2 - \tilde{\cY}_2 \partial \tilde{\cY}_1\,, \\ \hat{\tilde{\cT}}
&= \tilde{\cY}_1 \partial \tilde{\cT}_2 -\half \partial \tilde{\cY}_1 \tilde{\cT}_2
-(1\leftrightarrow 2) +{\rm c.c.}\,. \\
      \end{aligned} \right.
\end{equation} 
In the basis \eqref{conformal basis riemann},
the commutation relations become
\begin{equation}
  \begin{split} &[ \tilde{\cY}_m, \tilde{\cY}_n ] = (m-n) \tilde{\cY}_{m+n} , \quad [
    \tilde{\bar\cY}_m, \tilde{\bar\cY}_n ] = (m-n)\tilde{\bar\cY}_{m+n}, \\ &[\tilde{\cY}_m ,
\mathscr{\tilde T}_{k,l} ] = ( \frac{1}{2} m - k)
\mathscr{\tilde T}_{m+k,l}, \quad [\tilde{\bar\cY}_m , \mathscr{\tilde T}_{k,l} ] =
(\frac{1}{2} m - l) \mathscr{\tilde T}_{k,m+l} \\ &[ \tilde{\cY}_m ,
\tilde{\bar\cY}_n ] = 0 = [ \mathscr{\tilde T}_{k,l} ,
\mathscr{\tilde T}_{r,s}] .
\end{split} \label{commutation in conformal basis}
\end{equation}

\subsection{Coadjoint representation of the algebra}
\label{sec:coadj-repr-algebra-1}

The coadjoint representation in the basis \eqref{eq:59} may be obtained from the
structure constants of the algebra contained in \eqref{commutation in conformal
  basis} using \eqref{eq:3}. Alternatively, it can be derived using the results
of subsection \ref{sec:algebra-1} together with the explicit expressions of the
generators and their duals \eqref{eq:59}, and also from \eqref{eq:63}.
Explicitly,
\begin{equation}
  \begin{split}
    &{\rm ad}^*_{\tilde{\cY}_m}\tilde{\cY}^n_* = (-2m+n)\tilde{\cY}^{n-m}_*,
    \quad {\rm ad}^*_{\tilde{\bar\cY}_m} \tilde{\bar\cY}^n_*  = (-2m+n)
    \tilde{\bar\cY}_*^{n-m}, \\ &{\rm ad}^*_{\tilde{\cY}_m}\mathscr{\tilde T}^{k,l}_* = (-
    \frac{3}{2} m +k ) \mathscr{\tilde T}^{k-m,l}_* , \quad {\rm ad}^*_{\tilde{\bar\cY}_m} 
    \mathscr{\tilde T}^{k,l}_* = ( -\frac{3}{2} m + l)
    \mathscr{\tilde T}^{k,l-m}_* , \\ &{\rm ad}^*_{\mathscr{\tilde T}_{k,l}}  \mathscr{\tilde T}^{r,s}_*
    = (\frac{r-3k}{2}) \delta^s_l\tilde{\cY}_*^{r-k} + ( \frac{s-
      3 l}{2} ) \delta^r_k\tilde{\bar\cY}^{s-l}_* , \\ &{\rm ad}^*_{\tilde{\cY}_m}
    \tilde{\bar\cY}^n_*  = 0 = {\rm ad}^*_{\tilde{\bar\cY}_m} \tilde{\cY}^n_*  , \quad
    {\rm ad}^*_{\mathscr{\tilde T}_{k,l}} \tilde{\cY}^m_* = 0 = {\rm ad}^*_{\mathscr{\tilde T}_{k,l}} 
    \tilde{\bar\cY}^m_* .\label{coadjoint in conformal basis}
\end{split}
\end{equation}

\section{Comments on the cylinder}
\label{sec:realization-cylinder}

The mapping from the punctured plane to the vertical cylinder is standard in the
context of conformal field theory. It is defined through
\begin{equation}
  \label{eq:68}
  z = e^{-i\frac{2\pi}{L_1} w}, \quad w=w_1+iw_2,\quad w_1\sim w_1+L_1.
\end{equation}
According to \eqref{eq:75a}, conformal fields on the cylinder are related to
those on the punctured plane through
\begin{equation}
  \label{eq:69}
   \phi^{C_V}_{h,\bar h}(w,\bar w)=\big(-i\frac{2\pi}{L_1}z\big)^{h}\big(i\frac{2\pi}{L_1}\bar z\big)^{\bar h}\
   \phi_{h,\bar h}(z,\bar z). 
\end{equation}
When naively substituting the expansion adapted to the punctured plane
\eqref{Laurent series}, the associated expansion on the cylinder is
\begin{equation}
  \label{eq:71}
   \phi^{C_V}_{h,\bar h}(w,\bar w)=
  \sum_{k,l}\ a_{k,l}\ {}_{h,\bar h} Z^{C_V}_{k,l},\quad {}_{h,\bar h} Z^{C_V}_{k,l}
  =i^{\bar h-h}\big(\frac{2\pi}{L_1}\big)^{h+\bar h} e^{i\frac{2\pi }{L_1}k w}\ e^{-i\frac{2\pi }{L_1}l \bar w}, 
\end{equation}
with $k,l$ semi-integer when $h,\bar h$ are semi-integer. As usual for
Neveu-Schwarz boundary conditions, it follows that for half-integer conformal
weights, holomorphic or anti-holomorphic fields on the cylinder are
anti-periodic.

The generators \eqref{conformal basis riemann} of the $\mathfrak{bms}_4$ algebra
become
\begin{equation}
  \label{eq:75}
  \cY^{C_V}_m=i\big(\frac{2\pi}{L_1}\big)^{-1}e^{i\frac{2\pi }{L_1}m w},\ {\bar\cY}^{C_V}_m
  =-i\big(\frac{2\pi}{L_1}\big)^{-1}e^{-i\frac{2\pi }{L_1}m \bar w},\
  {\cT}^{C_V}_{k,l}=\big(\frac{2\pi}{L_1}\big)^{-1}e^{i\frac{2\pi }{L_1}k w}e^{-i\frac{2\pi }{L_1}l \bar w}, 
\end{equation}
while those of the coadjoint representation \eqref{eq:59} become 
\begin{equation}
  \label{eq:76}
  \cY^{m}_{C_V*}=i\big(\frac{2\pi}{L_1}\big)^{3}e^{i\frac{2\pi}{L_1} m \bar w},\ {\bar\cY}^m_{C_V*}
  =-i\big(\frac{2\pi}{L_1}\big)^{3}e^{-i\frac{2\pi }{L_1}m w},\
  {\cT}^{k,l}_{C_V*}=\big(\frac{2\pi}{L_1}\big)^{3}e^{-i\frac{2\pi }{L_1}l w}e^{i\frac{2\pi }{L_1}k \bar w}. 
\end{equation}
By construction, the commutation relations of the elements in \eqref{eq:75} are
unchanged: they are obtained from \eqref{commutation in conformal basis} by
adding a superscript $C_V$ to the generators.

For the coadjoint representation, matters are more subtle. It remains true that
the vector space generated by the elements of \eqref{eq:76} is a representation
of $\mathfrak{bms}_4$ in the sense of Remark (v) of section \ref{sec:algebra-1},
which is explicitly given by adding a superscript, respectively subscript, $C_V$
to the generators of \eqref{coadjoint in conformal basis}. This is not however
the coadjoint representation since there are issues with the pairing on the
infinite cylinder. Indeed,
\begin{multline}
  \label{eq:81}
  \langle \psi^{C_V}_{-\bar h+1,-h+1},\phi^{C_V}_{h,\bar h}
  \rangle_{C_V}\equiv\frac{1}{8\pi^2}\int idw\wedge d\bar w\ \big[\overline{
    \psi^{C_V}_{-\bar h+1,-h+1}}\
  \phi^{C_V}_{h,\bar h}\big]\\
  =\frac{1}{4\pi^2}\int_0^{L_1}dw_1\int^\infty_{-\infty}dw_2\ \big[\overline{
    \psi^{C_V}_{-\bar h+1,-h+1}}\
\phi^{C_V}_{h,\bar h}\big],
\end{multline}
and in particular,
\begin{multline}
  \label{eq:82}
  \langle {}_{-\bar h+1,-h+1} Z^{C_V}_{k',l'},{}_{h,\bar h} Z^{C_V}_{k,l} \rangle_{C_V}=
  \frac{1}{L_1}\delta^{k+l'}_m\delta^{k'+l}_m\quad\int^{+\infty}
  _{-\infty}dw_2 \ e^{-\frac{4\pi}{L_1}w_2m}\\
  =\frac{1}{2}\delta^{k+l'}_m\delta^{k'+l}_m\quad \int^{+\infty}_{-\infty} \frac{d\kappa}{2\pi}\, e^{\kappa m}.
\end{multline}
The remaining integral does not impose $m=0$, as in \eqref{eq:60} in the context
of the coadjoint representation on the punctured plane. For the infinite
vertical cylinder, the functions
\begin{equation}
{}_{h,\bar h} Z^{C_V}_{k,l}=i^{\bar h-h}\big(\frac{2\pi}{L_1}\big)^{h+\bar h}e^{i\frac{2\pi
  }{L_1}(k-l) w_1}\ e^{-\frac{2\pi }{L_1}(k+l)w_2},\label{eq:90}
\end{equation}
are not appropriate for expansions.
One should rather use
\begin{equation}
  \label{eq:83}
  \phi^{C_V}_{h,\bar h}(w,\bar w)= \sum_{m}\int^{+\infty}_{-\infty} d\kappa\ a_{m}(\kappa) Z^{C_V}_m(\kappa),\quad Z^{C_V}_m(\kappa)=i^{\bar h-h}
  \big(\frac{2\pi}{L_1}\big)^{h+\bar h}
  e^{i\frac{2\pi}{L_1}m w_1}e^{i\frac{2\pi}{L_1}\kappa w_2},
\end{equation}
which satisfy
\begin{equation}
  \label{eq:101}
  \langle {}_{-\bar h+1,-h+1} Z^{C_V}_{m'}(\kappa'),{}_{h,\bar h} Z^{C_V}_{m}(\kappa) \rangle_{C_V}=\delta^{m'}_{m} \delta(\kappa'-\kappa).
\end{equation}

\section{Identification in non-radiative asymptotically flat spacetimes}
\label{sec:embedd-into-asympt}

We limit ourselves in this section to the case of the sphere. Asymptotically
flat space-times in the Newman-Penrose-Unti sense are for instance defined in
\cite{Penrose:1986}, end of section 9.8. Here we consider the case with the
Maxwell field turned off, $\varphi_1=0=\varphi_2$. Non-radiative spacetimes
correspond to the subset of solutions with $u$-independent asymptotic part of
the shear,
\begin{equation}
\partial_u\sigma^0=0,\label{eq:45}
\end{equation}
(as well as its complex conjugate and all higher order $u$ derivatives), so that
the news and also $\Psi^0_3,\Psi^0_4$ vanish. It follows that
\begin{equation}
\Psi^0_2-\xbar
\Psi^0_2=\xbar\eth^2\sigma^0-\eth^2\xbar\sigma^0,\label{eq:46}
\end{equation}
while the evolution equations imply that 
\begin{equation}
\partial_u\Psi^0_2=0,\quad \Psi^0_1=\Psi^0_1(\xi,\bar\xi)+u\eth \Psi^0_2. \label{eq:47}
\end{equation}
Such non-radiative space-times are completely characterized by specifying, at
the cut $u=0$ of $\scri^+$, the free data
\begin{equation}
  \label{eq:50}
   \Psi^0_2+\xbar\Psi^0_2,\Psi^0_1,\sigma^0,
\end{equation}
together with the different orders $\Psi^n_0$ in a $1/r$ expansion of $\Psi_0$, 
\begin{equation}
\Psi_0=\sum_{n\geq 0}\Psi^n_0(\xi,\bar\xi)\,r^{-5-n}\label{eq:80}
\end{equation}
Besides the linear dependence $u$ dependence of $\Psi^0_1$ in \eqref{eq:47},
there are also evolution equations that govern the $u$-dependence of $\Psi^n_0$
which do not concern us here.

The transformation of this data under BMS symmetries has been worked out in
different ways and under various assumptions in
\cite{Barnich:2013oba,Barnich:2013axa,Barnich:2016lyg,Barnich:2019vzx}\footnote{The
  arxiv version of the last reference is preferable to the published one on
  account of typesetting issues in the latter.} In the case of the sphere,
$P=P_S$ and the scalar curvature is $R_S=2$. Furthermore, the solutions
$\cY,\xbar\cY$ to the conformal Killing equation on the sphere are given by
\eqref{Y decomposition} and \eqref{eq:22}. This implies in particular that
\begin{equation}
  \label{eq:53}
  \eth^3\cY=0\quad \eth R_S=0, 
\end{equation}
together with the complex conjugate relations\footnote{Note that in the
  considerations below the value of $R_S=2$ on the sphere is never needed, only
  the second of \eqref{eq:53} is used}. In the non-radiating case and at $u=0$,
the infinitesimal transformations reduce to\footnote{Up to a conventional
  overall sign that we have changed here.}
\begin{equation}
  \label{eq:51}
  \begin{split}
    \delta_s \Psi^0_2 &=[\cY \eth +\xbar \cY
    \xbar\eth + \frac32 \eth\cY + \frac32\xbar\eth\xbar\cY]\Psi^0_2 ,\\
    \delta_s \Psi^0_1&=[\cY \eth +\xbar \cY \xbar\eth + 2 \eth\cY +
    \xbar\eth\xbar\cY]\Psi^0_1 +
    \cT\eth\Psi_2^0 + 3 \eth \cT \Psi^0_2 ,\\
    \delta_s \sigma^0&=[\cY \eth +\xbar \cY \xbar\eth + \frac32 \eth\cY -
    \frac12 \xbar\eth\xbar\cY ]\sigma^0 - \eth^2 \cT,\\
    \delta_s \Psi^0_0&=[\cY \eth +\xbar \cY \xbar\eth + \frac52 \eth\cY +
    \frac12\xbar\eth\xbar\cY ]\Psi^0_0 + \cT\eth\Psi_1^0 + 3\cT\sigma^0\Psi_2^0 + 4
    \eth \cT  \Psi^0_1 ,\\
    \delta_s \Psi^1_0 &=\big[\cY \eth +\xbar \cY \xbar\eth + 3 \eth\cY +
    \xbar\eth\xbar\cY\big]\Psi^1_0 - \overline{\eth}\big[5 \eth \cT \Psi^0_0 + \cT
    \eth \Psi^0_0 + 4 \cT \Psi^0_1 \sigma^0 \big].
\end{split}
\end{equation}

There are increasingly complicated transformations laws for the higher
$\Psi^n_0$, $n\geq 2$, that are not relevant for our purpose here. 

When expressing the first two of the equations in \eqref{eq:51} in terms of the
free data by taking the constraint \eqref{eq:46} into account, one finds
(trivially) that
\begin{equation}
  \label{eq:24}
  \delta_s(\Psi^0_2+\xbar\Psi^0_2)=[\cY \eth +\xbar \cY
  \xbar\eth + \frac32 \eth\cY + \frac32\xbar\eth\xbar\cY](\Psi^0_2+\xbar\Psi^0_2),
\end{equation}
and
\begin{multline}
  \label{eq:42}
  \delta_s\Psi^0_1=[\cY \eth +\xbar \cY \xbar\eth + 2 \eth\cY +
  \xbar\eth\xbar\cY]\Psi^0_1 +
  \frac 12 \cT\eth(\Psi_2^0+\xbar\Psi^0_2+\xbar\eth^2\sigma^0-\eth^2\xbar\sigma^0)
  \\ + \frac 32 \eth \cT (\Psi_2^0+\xbar\Psi^0_2+\xbar\eth^2\sigma^0-\eth^2\xbar\sigma^0). 
\end{multline}
Following the analysis in three dimensions, one fixes the normalization by
computing the surface charge algebra, directly related to linear super momentum
and angular momentum of the system. In the non-radiating case, this has been
discussed for instance in section 4.2 of \cite{Barnich:2013axa} (see also
\cite{Barnich:2019vzx}). Let us summarize the relevant part of those results in
the notation and conventions adopted here. Let
\begin{equation}
  \label{eq:43}
  f=\cT+\frac 12 u (\eth\cY+\xbar\eth\xbar\cY), 
\end{equation}
and consider the $2$-form
\begin{equation}
  \label{eq:48}
  J_s=\frac{i}{R^2}\big[(P_S\bar P_S)^{-1}\cJ^u_sd\xi\wedge d\bar\xi
  +P_S^{-1}\cJ_s^{\bar\xi}du\wedge d\xi-\bar P_S^{-1}\cJ_s^\xi du\wedge d\bar\xi ], 
\end{equation}
with $\cJ^\xi_s= \cJ_s$, $\cJ^{\bar\xi}=\xbar\cJ$ and 
\begin{equation}
  \label{eq:52}
  \begin{split}
    \cJ^u_s&=-\frac{1}{8\pi G} \big[(\Psi^0_2+\xbar\Psi^0_2)f+\Psi^0_{1\bar J}\cY+\xbar\Psi^0_{1\bar J}\xbar\cY\big],\\
    \cJ_s&=\frac{1}{8\pi G} \big[\Psi^0_2\cY+\frac 12 \eth\xbar\sigma^0(\eth\cY-\xbar\eth\xbar\cY)
    -\frac 12 \xbar\sigma^0 \eth(\eth\cY-\xbar\eth\xbar\cY)\big],\\
    \Psi^0_{1\bar J}&=\Psi^0_{1}+\sigma^0\eth\xbar\sigma^0+\frac{1}{2}\eth(\sigma^0\xbar\sigma^0).
  \end{split}
\end{equation}
The transformation law of $\Psi^0_{1\bar J}$ turns out to be
\begin{multline}
  \label{eq:56}
  \delta_s \Psi^0_{1\bar J}=[\cY \eth + 2 \eth\cY]\Psi^0_{1\bar J} + \xbar\eth(\xbar\cY \Psi^0_{1\bar J})+
  \frac{1}{2}\cT\eth (\Psi^0_2+\xbar\Psi^0_2)+ \frac{3}{2} \eth \cT
  (\Psi^0_2+\xbar\Psi^0_2)\\
  +\frac12 \xbar\eth (\cT\xbar\eth\eth
  \sigma^0-\xbar\eth\cT\eth\sigma^0+3\eth\cT\xbar\eth\sigma^0-3\xbar\eth\eth\cT\sigma^0-\frac{3}{2}R_S\cT\sigma^0)
  -\frac 12 \eth^3(\cT\xbar\sigma^0),
\end{multline}
where the terms on the second line are irrelevant when multiplied by $\cY$ and
integrated over the sphere (cf. Remark (i) in section
\ref{sec:spin-weight-spher}).

When taking the retarded time-dependence of $\Psi^0_1$ in \eqref{eq:47} and the
constraint \eqref{eq:46} into account, this 2-form is closed,
\begin{equation}
  \label{eq:54}
  dJ_s=0\iff \partial_u \cJ^u_s+\eth \cJ_s+\xbar\eth\xbar{\cJ_s}=0.
\end{equation}
Furthermore,
\begin{equation}
  \label{eq:55}
  \delta_{s_1}\cJ^u_{s_2}=-\cJ^u_{[s_1,s_2]}+\eth \cL_{s_2,s_1}+\xbar\eth\overline{\cL_{s_2,s_1}},
\end{equation}
where the concrete expression for $\cL_{s_2,s_1}$ is not needed here. This
transformation law is in line with \eqref{eq:9}. The charges defined by
\begin{equation}
  \label{eq:116}
  Q_s=\int_{S^2,u=u_0} J^u_s,
\end{equation}
with $u_0$ constant, are conserved in the sense that they do not depend on $u$
and
\begin{equation}
  \label{eq:117}
  \delta_{s_1} Q_{s_2}=-Q_{[s_1,s_2]}. 
\end{equation}

More precisely, the polynomial algebra $\cF$ generated by the free data
$\Psi^0_2+\xbar \Psi^0_2,\Psi^0_{1},\sigma^0,\xbar \Psi^0_{1},\xbar \sigma^0$
carries a representation $\delta_s$ of the BMS$_4$ algebra. It then follows from
the identification at $u=0$,
\begin{equation}
  \label{eq:118}
  Q_s=\langle ([J],[\bar J],\cP),(\cY,\bar\cY,\cT) \rangle,
\end{equation}
that the pre-moment map $\mu:\cF\to \mathfrak{bms}^*_4$ defined by
\begin{equation}
  \label{eq:119}
  \mu(-\frac{1}{2G}[\Psi^0_2+\xbar \Psi^0_2])
  =\cP,\quad \mu(-\frac{1}{2G}\Psi^0_{1\bar J})=[\bar\cJ],\quad  \mu(-\frac{1}{2G}\xbar{\Psi^0_{1\bar J}})=[\cJ], 
\end{equation}
is compatible with the representation, 
\begin{equation}
\mu\circ\delta_s={\rm ad}^*_s\circ \mu \label{eq:122}.
\end{equation}

The transformation law of the asymptotic part of the shear implies in
particular that 
\begin{equation}
  \label{eq:41}
  \delta_s (\xbar\eth^2\sigma^0-\eth^2\xbar\sigma^0)=[\cY \eth +\xbar \cY
  \xbar\eth + \frac32 \eth\cY + \frac32\xbar\eth\xbar\cY](\xbar\eth^2\sigma^0-\eth^2\xbar\sigma^0). 
\end{equation}
This means that constraining the asymptotic part of the shear to be electric
\cite{Newman1966},
\begin{equation}
  \label{eq:57}
  \xbar\eth^2\sigma^0_e=\eth^2\xbar\sigma^0_e,
\end{equation}
is a BMS invariant condition. In this case, the transformation law \eqref{eq:42}
simplifies and suggests
\begin{equation}
  \label{eq:103}
  \mu(-\frac{1}{2G}\Psi^0_1)=\bar\cJ\sim\Psi^0_1,\quad \mu(-\frac{1}{2G}\xbar \Psi^0_1)=\cJ. 
\end{equation}
That this is compatible with the previous identification can be seen as follows.
The electric condition is solved by a real field $\chi_e=\xbar\chi_e$ with the
same weights $s=0,w=1$ than $\cT$, 
\begin{equation}
  \label{eq:110}
  \sigma^0_e=\eth^2 \chi_e,\quad
  \xbar\sigma^0=\xbar\eth^2\chi_e,\quad\delta_s\chi_e
  =[\cY \eth +\xbar \cY \xbar\eth
  -\frac 12 \eth\cY -\frac 12
  \xbar\eth\xbar\cY]\chi_e-\cT+\sum_{j\leq 1,m}\lambda^{jm}{}_0Z_{j,m},
\end{equation}
where $\lambda^{jm}\in \mathbb R$. Inserting this solution into
$\Psi^0_{1\bar J}$ one finds that it indeed agrees with $\Psi^0_1$ up
to terms that are projected to zero by the map,
\begin{equation}
  \label{eq:113}
  \Psi^0_{1\bar J}=\Psi^0_1+\frac 12
  \xbar\eth(\eth^3\chi_e\xbar\eth\chi_e
  +3\eth^2\chi_e\eth\xbar\eth\chi_e
  -\frac 34 R_S \eth\chi_e\eth\chi_e)-\frac 14 \eth^3(\xbar\eth\chi_e\xbar\eth\chi_e).
\end{equation}

Relevant formulas for the group can be found in \cite{Barnich:2016lyg} and will
not be repeated here. On the punctured plane, in order to have room for the Witt
algebra, one cannot limit oneself to non-radiative spacetimes since turning off
the news requires $\partial^3\tilde\cY=0=\xbar\partial^3 \xbar{\tilde\cY}$. In the
presence of news, currents are no longer conserved. Current algebra is broken
both by flux terms and by a field dependent central extension discussed in more
details in \cite{Barnich:2017ubf}. In this case, the last term in \eqref{eq:56}
is no longer trivial and becomes the associated (field-dependent) Souriau
cocyle.

\section{Discussion and perspectives}
\label{sec:discussion}

For the generalized BMS group on the sphere introduced in
\cite{Campiglia:2014yka} (see also
\cite{Compere:2018ylh,Flanagan:2019vbl,Campiglia:2020qvc,Compere:2020lrt}
for further considerations), the coadjoint representation is obtained
from the approach developed here simply by removing the conformal
Killing equation on infinitesimal superrotations
$\xbar\eth \cY=0=\eth\xbar\cY$ and the associated equivalence
relations on super-angular momentum $\cJ, \xbar\cJ$. All fields should
then simply be expanded in terms of spin-weighted spherical harmonics
according to their weights.

A detailed recent study of the coadjoint representation of closely related
semi-direct product groups involving diffeomeorphisms on the sphere, along the
lines of our analysis in three dimensions \cite{Barnich:2014kra,Barnich:2015uva}
(see also \cite{Oblak:2016eij} for a review), has recently appeared in
\cite{Donnelly:2020xgu}.

For the $\mathfrak{bms}_4$ algebra on the punctured plane, a discussion of
central extensions can be found in \cite{Barnich:2011ct}, whereas deformations
have been studied in detail in \cite{Safari:2019zmc}.

After having set up the basics in this paper, the next steps are to
classify the coadjoint orbits, to re-discuss unitary irreducible
representations
\cite{Cantoni1966,McCarthy1972a,McCarthy1973,McCarthy1973a,Girardello:1974sq,McCarthy1975}
from the viewpoint of the orbit method \cite{A.A.Kirillov897} and to
construct the associated geometric actions
\cite{Alekseev:1988vx,Alekseev1989}, as in the three dimensional case
\cite{Barnich:2017jgw} (see \cite{Nguyen:2020hot} in this
context). One could also explore whether some aspects of positive
energy theorems for the Bondi mass
\cite{Israel1981,Ludvigsen:1981gf,Schon:1982re,Horowitz:1981uw,Horowitz1982,Ashtekar1982a,Reula1984}
might be understood from such a perspective, again as in three
dimensions \cite{Barnich:2014zoa}.

The most interesting question is to understand in detail how such
effective actions for the sector captured by the coadjoint
representation interacts with the radiative degrees of freedom, as
described in \cite{Ashtekar1981a,Ashtekar:1987tt} and more recently in
\cite{AdamoCasaliSkinner2014}, see also
\cite{Troessaert:2015nia,Wieland:2020gno} in this context.

Another more technical question is to extend the considerations in section
\ref{sec:embedd-into-asympt} to a full-fledged momentum map at null infinity, as
recently constructed at spatial infinity
\cite{Troessaert:2017jcm,Henneaux:2018cst}, by starting from \cite{Torre:1985rw}
and also \cite{Oliveri:2019gvm}.

In the case of celestial scattering amplitudes and soft theorems, the
relevant surface is neither the (Riemann) sphere nor the plane, but
rather two Riemann spheres with punctures related by an antipodal
map. On each of these surfaces, the superrotation part of the extended
algebra is given by the Witt algebra only if there are two
particles/punctures. In this context, complementary aspects of
the BMS group have been discussed in
\cite{Strominger:2013jfa,He:2014laa,Conde:2016rom,Cheung:2016iub,Banerjee:2018fgd,%
Distler:2018rwu,Donnay:2018neh,Stieberger2019,Adamo:2019ipt,Fotopoulos:2019tpe,Fotopoulos2020,%
Fan:2020xjj,Donnay:2020guq,Donnay:2020fof,Gonzalez:2020tpi,Banerjee:2020kaa,%
Banerjee:2020zlg,Himwich:2020rro,Atanasov:2021oyu}. 

In this exposition here, we have followed the general relativity route
from the sphere to the punctured plane. In conformal field theory, one
travels in the opposite direction. For more punctures, the appropriate
algebra should presumably be Krichever-Novikov algebras
\cite{Krichever:1987rm,Krichever1987}.

\section*{Acknowledgments}
\label{sec:acknowledgements}

\addcontentsline{toc}{section}{Acknowledgments}

The authors thank P.~Mao for collaboration on related questions and
are grateful to L.~Ciambelli, Y.~Herfray, M.~Petropolous, B.~Oblak,
C.~Marteau and L.~Szabados for insightful discussions. The work of
G.B.~is supported by the F.R.S.-FNRS, Belgium, convention FRFC PDR
T.1025.14 and convention IISN 4.4514.08. R.R.~is supported by the
Austrian Science Fund (FWF), project P 32581-N.

\appendix

\section{Spin-weighted spherical harmonics}
\label{SWSH}

We follow the conventions of \cite{Penrose:1984}, section
4.15. Instructive alternative presentations and perspectives can be
found in \cite{Wu:1976ge}, \cite{Thorne1980},
\cite{Eastwood:1982aa}, \cite{A.S.Galperin13674}, section 1.10,
\cite{Beyer:2013loa}, \cite{Boyle:2016tjj}.
 
Let
\begin{equation}
  \begin{pmatrix}
    \alpha &\beta \\ \gamma &\delta
  \end{pmatrix} = \frac{i}{\sqrt{1+ \xi \bar{\xi}}}
  \begin{pmatrix} -1 &
\xi \\ \bar{\xi} & 1
\end{pmatrix}.
\label{def Z 2}
\end{equation}
Let also $s$ be integer or half-integer, $j\pm s\in \mathbb N$,
$j\pm m\in \mathbb N$, $|m| \le j$, $|s| \le j$, and consider
\begin{equation}
  _sZ_{j,m} = \sum_r
\frac{(j+m)!(j-m)!(j+s)!(j-s)!\alpha^r \beta^{j-m-r}\gamma^{j+s-r}
\delta^{r+m-s}}{(2j)!r!(j-m-r)!(j+s-r)!(r+m-s)!},
\label{def Z}
\end{equation}
where the summation extends over integer values of $r$ in the range $\max(0, 
s-m) \le r \le \min(j-m, j+s)$.

The spin-weighted spherical harmonics $_sY_{j,m}$ are then defined by
\begin{equation}
  _sY_{j,m} = (-1)^{j+m} ~_sZ_{j,m}
\sqrt{\frac{(2j+1)!(2j)!}{4 \pi (j+s)!(j-s)!(j+m)!(j-m)!}}.
\label{Y in term of Z}
\end{equation}
For $s=0$, one recovers the usual spherical harmonics functions, i.e., $_0
Y_{j,m}$. The following properties hold:
\begin{itemize}
\item For each $s$, the ${}_s Z_{j,m}$ form an orthogonal basis for the
  spin-weighted scalars $\eta^{s}$ on the sphere with
\begin{equation}
  \langle {}_{s}Z_{j',m'}, {}_sZ_{j,m}\rangle =
  \frac{(j+s)!(j-s)!(j+m)!(j-m)!}{(2j+1)!(2j)!}\delta_{jj'} \delta_{mm'},\label{normZ} 
\end{equation}
for the pairing \eqref{eq60}. The ${}_s Y_{j,m}$ form an orthonormal
basis with
\begin{equation}
  4 \pi \langle {}_{s}Y_{j',m'}, {}_sY_{j,m}\rangle =
  \delta_{jj'} \delta_{mm'}.
\end{equation}

\item The behavior under complex conjugation is 
  \begin{equation}
    \overline{_sZ_{j,m}} = (-1)^{m+s} ~_{-s}Z_{j,-m},\quad
    \overline{_sY_{j,m}} = (-1)^{3m+s} ~_{-s}Y_{j,-m}. 
\end{equation}

\item The action of the operators $\eth$ and $\bar{\eth}$ defined in \eqref{eth
    operators sphere} is explicitly given by
\begin{equation}
\eth\ {}_s Z_{j,m} = - \left( \frac{j-s}{R\sqrt{2}}
\right) {}_{s+1} Z_{j,m}, \quad \bar{\eth}\ {}_sZ_{j,m} = \left(
\frac{j+s}{R\sqrt{2}} \right) {}_{s-1} Z_{j,m}.
\label{eth acting on Z}
\end{equation}
\begin{equation}
  \begin{split}
\eth\ _sY_{j,m} &= - \sqrt{ \frac{(j+s+1)(j-s)}{2R^2}}
{}_{s+1}Y_{j,m}, \\ \bar{\eth}\ {}_sY_{j,m} &= \sqrt{
\frac{(j-s+1)(j+s)}{2R^2} } {}_{s-1}Y_{j,m}.
\end{split}
\end{equation}

\item The ${}_sZ_{j,m}$ and ${}_sY_{j,m}$ are eigenfunctions of the
operator $\bar{\eth} \eth$:
\begin{equation}
  \bar{\eth} \eth _sZ_{j,m} = -(j+s+1)(j-s) \frac{1}{2}
  ~_sZ_{j,m} ,\quad \bar{\eth} \eth _sY_{j,m} = -(j+s+1)(j-s) \frac{1}{2}
  ~_sY_{j,m} .
\end{equation}
\item Products of spin-weighted spherical harmonics can be decomposed
  as
  \begin{equation}
\begin{split} _{s_1}Z_{j_1,m_1}~_{s_2}Z_{j_2,m_2} &= \sqrt{\frac{(j_1+
s_1)!(j_1- s_1)!(j_1+ m_1)!(j_1 - m_1)!}{(2j_1)! (2j_1)!}} \\ &\times
\sqrt{\frac{(j_2+ s_2)!(j_2- s_2)!(j_2+ m_2)!(j_2 - m_2)!}{(2j_2)!
(2j_2)!}} \\ &\times \sum_j (-1)^{j_1 + j_2 +j} ~_{(s_1 + s_2)}Z_{j,
(m_1 + m_2)} \\ &\times
\sqrt{\frac{(2j)!(2j)!}{(j+s_1+s_2)!(j-s_1-s_2)! (j+m_1+
m_2)!(j-m_1-m_2)!}} \\ & \times \langle j_1, s_1 ; j_2, s_2 | j, (s_1
+ s_2) \rangle ~ \langle j_1, m_1; j_2, m_2 | j, (m_1 + m_2) \rangle,
\end{split}
\label{product of SWSH}
\end{equation}
or
\begin{equation}
\begin{split} _{s_1}Y_{j_1,m_1}~_{s_2}Y_{j_2,m_2} =& \sum_j
\sqrt{\frac{(2j_1+ 1)(2j_2+1)}{4 \pi (2j+1)}} ~_{(s_1 + s_2)}Y_{j,
(m_1 + m_2)} \\ & \times \langle j_1, s_1 ; j_2, s_2 | j, (s_1 + s_2)
\rangle ~ \langle j_1, m_1; j_2, m_2 | j, (m_1 + m_2) \rangle ,
\end{split}
\end{equation}
where the summation extends over integer values of $j$ in the range
$\max(|j_1-j_2|, |s_1 + s_2|, |m_1+m_2|) \le j \le j_1 + j_2$, and where
$\langle j_1, m_1; j_2, m_2 | j, (m_1 + m_2) \rangle$ is a Clebsch-Gordan
coefficient of the rotation group (see e.g.~\cite{Held1970,Beyer:2013loa} in
this context). 

\end{itemize}

\providecommand{\href}[2]{#2}\begingroup\raggedright\endgroup

%\bibliography{C:/Users/gbarn/Dropbox/Literature/master}
%\bibliography{/mnt/c/Users/gbarn/Dropbox/Literature/master}
%\bibliography{master}
%\bibliography{masterbiblio}

\end{document}